\documentclass[10pt,journal,compsoc]{IEEEtran}
\ifCLASSOPTIONcompsoc
  \usepackage[nocompress]{cite}
\else
  \usepackage{cite}
\fi
\ifCLASSINFOpdf
\else
\fi
\usepackage[normalem]{ulem}
\usepackage{caption}

\usepackage{cite}
\usepackage{amsmath,amssymb,amsfonts}
\usepackage{tcolorbox}
\usepackage{algorithmic}
\usepackage{graphicx}
\usepackage{balance}
\usepackage{textcomp}
\usepackage{xcolor}
\usepackage{url}
\usepackage{multirow}
\usepackage[normalem]{ulem}

\usepackage{hyperref}
\hypersetup{
    colorlinks=true,
    linkcolor=blue,
    filecolor=blue,      
    urlcolor=blue,
    citecolor=blue,
    pdfpagemode=FullScreen,
    }
    
\usepackage[framemethod=tikz]{mdframed}

\usepackage{tikz}

\mdfdefinestyle{mpdframe}{
    frametitlebackgroundcolor   =black!15,
    frametitlerule              =true,
    roundcorner                 =5pt,
    middlelinewidth             =1pt,
    innermargin                 =0.2cm,
    outermargin                 =0.2cm,
    innerleftmargin             =0.2cm,
    innerrightmargin            =0.2cm,
    innertopmargin              =0.2cm,
    innerbottommargin           =0.2cm
}
\newcommand{\toolname}[0]{MiDas}
\DeclareUnicodeCharacter{2212}{-}

\begin{document}
\title{Multi-Granularity Detector for Vulnerability Fixes}

\author{Truong Giang Nguyen,
        Thanh Le-Cong,
         Hong Jin Kang,
         Ratnadira Widyasari, 
         Chengran Yang,
         Zhipeng Zhao,
         Bowen Xu,
         Jiayuan Zhou,
         Xin Xia,
         Ahmed E. Hassan,
         Xuan-Bach D. Le,
         and David Lo
\IEEEcompsocitemizethanks{\IEEEcompsocthanksitem Truong Giang Nguyen, Thanh Le-Cong, Hong Jin Kang, Ratnadira Widyasari, Chengran Yang, Zhipeng Zhao, Bowen Xu, David Lo are with the School of Computing and Information Systems, Singapore Management University, Singapore.\protect\\
E-mail:\{gtnguyen, tlecong, hjkang.2018, ratnadiraw.2020, cryang, zpzhao, bowenxu.2017, davidlo\} @smu.edu.sg.
\IEEEcompsocthanksitem Jiayuan Zhou and Xin Xia are with the Software Engineering Application Technology Lab, Huawei, China.\protect\\
E-mail: jiayuan.zhou1@huawei.com, xin.xia@acm.org
\IEEEcompsocthanksitem Ahmed E. Hassan is with School of Computing, Queen's University, Canada \\
E-mail: ahmed@cs.queensu.ca
\IEEEcompsocthanksitem Xuan-Bach D. Le is  with School of Computing and Information Systems, The University of Melbourne, Australia \protect\\
E-mail: bach.le@unimelb.edu.au
\IEEEcompsocthanksitem Bowen Xu is the corresponding author.
}% <-this % stops an unwanted space
% \thanks{Manuscript received April 19, 2005; revised August 26, 2015.}
}

\IEEEtitleabstractindextext{%
\begin{abstract}
With the increasing reliance on Open Source Software, users are exposed to third-party library vulnerabilities. Software Composition Analysis (SCA) tools have been created to alert users of such vulnerabilities. SCA requires the identification of vulnerability-fixing commits. Prior works have proposed methods that can automatically identify such vulnerability-fixing commits. However, identifying such commits is highly challenging, as only a very small minority of commits are vulnerability fixing. Moreover, code changes can be noisy and difficult to analyze. We observe that noise can occur at different levels of detail, making it challenging to detect vulnerability fixes accurately. 

To address these challenges and boost the effectiveness of prior works, we propose \toolname{} (Multi-Granularity Detector for Vulnerability Fixes). Unique from prior works, \toolname{} constructs different neural networks for each level of code change granularity, corresponding to commit-level, file-level, hunk-level, and line-level, following their natural organization. It then utilizes an ensemble model that combines all base models to generate the final prediction. This design allows \toolname{} to better handle the noisy and highly imbalanced nature of vulnerability-fixing commit data. Additionally, to reduce the human effort required to inspect code changes, we have designed an effort-aware adjustment for \toolname{}'s outputs based on commit length. The evaluation results demonstrate that \toolname{} outperforms the current state-of-the-art baseline in terms of AUC by 4.9\% and 13.7\% on Java and Python-based datasets, respectively. Furthermore, in terms of two effort-aware metrics, EffortCost@L and Popt@L, \toolname{} also outperforms the state-of-the-art baseline, achieving improvements of up to 28.2\% and 15.9\% on Java, and 60\% and 51.4\% on Python, respectively.
\end{abstract}

}

\maketitle

\IEEEpeerreviewmaketitle

\section{Introduction} \label{sec:introduction}

Nowadays, software projects are more and more reliant on third-party libraries, therefore exposed to these libraries' vulnerabilities.
As an example, a vast number of applications and cloud services that use Log4J, including Steam, Apple iCloud, and Minecraft, are affected by the Log4Shell vulnerability~\cite{log4j_vuln_1, log4j_vuln_2}. Log4Shell targets Log4J, one of the most popular Java libraries for logging messages and errors in the Java ecosystem. By logging a URI that points to a potentially untrusted Java class, attackers trick the client applications into executing malicious code. 

To avoid similar attacks, there has been increasing attention to addressing the growing problem of vulnerabilities propagated through libraries in a software ecosystem~\cite{liu2022demystifying, ponta2020detection, decan2018impact}. As developers are slow in updating their dependencies~\cite{gonzalez2017technical, ihara2017understanding, kula2018developers, shu2017study, chinthanet2021lags, zerouali2021multi}, tools have been developed to alert users of library vulnerabilities that may affect their applications~\cite{owasp_dependency_check, kang2022test, imtiaz2022open,pan2022automated}. For example, the Open Web Application Security Project (OWASP\footnote{\url{https://owasp.org/}}) foundation developed Dependency-Check \cite{owasp_dependency_check}, a tool that alerts users of publicly disclosed vulnerabilities within a project’s dependencies.

These tools, which are referred to as Software Component Analysis~\cite{zhou2017automated}, rely on databases of publicly disclosed vulnerabilities. Unfortunately, there is often a gap between the time a vulnerability is fixed and the time that a vulnerability is publicly disclosed ~\cite{imtiaz2022open}, e.g., the inclusion of the vulnerability in the National Vulnerability Database (NVD). For example, the fix for Log4Shell was pushed four days before its public disclosure. This gap of time creates a window of opportunity for attacker to develop an exploit before the vulnerability is even known. If a vulnerability is unknown, tools cannot be developed to detect it. To address this problem, previous studies~\cite{sabetta2018practical, nguyen2022hermes, nguyen2022vulcurator, zhoufinding} have propose tool to automatically detect security-relevant changes (i.e., vulnerability-fixing commits) that are not yet disclosed in open source projects.

Automatic identification of vulnerability-fixing commits has been used in many security companies such as Huawei, Veracode, Mend, and Snyk to monitored potential security issues from commits and other artifacts to provide users early warning of unpublished vulnerabilities ~\cite{pan2022automated,zhou2017automated, snyx_monitoring, whitesource_monitoring, james_insecure_2012}. It also can assist the security researchers in maintaining and updating the vulnerabilities database, such as National Vulnerability Database (NVD). Moreover, identifying vulnerability-fixing commits can enable applications such as hot patch generation and deployment~\cite{duan2019automating} and patch presence testing~\cite{dai2020bscout}. As substantial human effort is required to identify vulnerability-fixing commits manually, automated approaches to detect them are worth investigating. For example, a dataset of 1,282 vulnerability-fixing commits constructed in prior work required approximately four years to be manually curated~\cite{ponta2019manually}. Consequently, security companies have invested in building and deploying automated approaches to identify vulnerability-fixing commits to enhance IT supply chain security~\cite{zhou2017automated, sabetta2018practical, zhoufinding,zhoucolefunda}.

To address this problem, previous works~\cite{zhou2017automated, sabetta2018practical, chen2020machine, ramsauer2020sound, nguyen2022hermes, wu2022enhancing} leverage related resources of commits such as commit messages or issue reports to classify commits.
Unfortunately, in accordance with the good practice of coordinated vulnerability disclosure~\cite{vulnerability_disclosure, householder2017cert}, these resources should not mention any security-related information to fix vulnerabilities without exposing their existence before public disclosure of the vulnerability. 
Hence, detecting vulnerabilities and their corresponding fixes with the use of natural language resources such as commit messages or issue reports may be impractical.

Identifying vulnerability-fixing commits based on code-changes alone is an inevitable choice. However, traditional code analyses are not suitable for this task due to two main reasons: (1) most of these techniques cannot be applied to partial code, i.e. code changes in a commit~\cite{wu2022enhancing}, and (2) they require hand-crafted specifications or heuristics, which can be challenging and time-consuming to create~\cite{pradel2021neural}. An alternative solution is to use deep-learning-based analysis techniques, which can handle fuzzy inputs, including natural language integrated into code (e.g., meaning of variables' names \cite{pradel2018deepbugs}), and hidden patterns. These techniques have been shown to outperform traditional code analysis methods in various tasks, such as type inference~\cite{jesse2021learning, kazerounian2021simtyper}, fault localization~\cite{li2019deepfl, li2021fault, nguyen2022ffl} and program repair~\cite{chen2019sequencer, li2022dear, lin2022context, zhou2023patchzero, lutellier2020coconut}. Inspired by this success, Zhou et al. proposed VulFixMiner~\cite{zhoufinding}, which utilizes CodeBERT to automatically represent code changes and extract features for identifying vulnerability-fixing commits. Their empirical evaluation demonstrated that VulFixMiner can accurately identify 49\% of vulnerability-fixing commits with a minimal effort, inspecting only 5\% of the total lines of codes.

Although VulFixMiner has achieved positive performance, we found that there are aspects that are worth further investigation.
Commits could be tangled; a commit may contain changes related to different purposes, such as implementing new features and refactoring code~\cite{barnett2015helping}. In a tangled vulnerability-fixing commit, irrelevant changes may contribute to noise. The high noise may pose a challenge to a machine learning classifier. From our observations on real-world vulnerability-fixing commits, as illustrated in Section~\ref{sec:motivating}, noise can be presented at different levels of granularity, such as the file level, hunk level, or line level. 
Besides, the dataset of vulnerability-fixing commits is highly imbalanced, mainly because there are significantly fewer vulnerability-fixing commits compared to non-vulnerability fixing commits in the same project. For example, the vulnerability-fixing commits only account for 0.34\% of all commits in the VulFixMiner test dataset~\cite{zhoufinding}. The high data imbalance also poses a challenge to a machine learning classifier.

To address the aforementioned issues, we present MiDas (\underline{M}ult\underline{i}-Granularity \underline{D}etector for Vulner\underline{a}bility Fixe\underline{s}), an approach that constructs different base models for each level of code change granularity, corresponding to commit-level, file-level, hunk-level, and line-level, following their natural organization and then use an ensemble model combining all base models to output the final prediction. The benefit of \toolname{} are three-fold. Firstly, decomposing code changes into different levels of granularity allows \toolname{} to utilize a suitable extractor for each level, as discussed in Section~\ref{sec:extractor}. Secondly, ensemble learning helps to reduce errors caused by noise. According to previous research~\cite{polikar2012ensemble}, individual classifiers tend to make different errors on each sample but typically agree on their correct classifications. Thus, by combining multiple classifiers, ensemble learning can reduce the impact of noise in the data by averaging out the error components. Thirdly, ensemble learning has been shown to be effective in addressing data imbalance problems, as demonstrated in previous studies~\cite{haixiang2017learning, khoshgoftaar2015ensemble, dong2020survey}.

\vspace{2mm}

\noindent \textbf{Contribution.} In this paper, we made the following contributions:
\begin{itemize}
    \item We propose \toolname{}, a deep learning model, which utilizes multiple levels of granularity of code changes, along with an effort-aware adjustment to detect vulnerability-fixing commits.
    \item We demonstrate that our approach outperforms the current state-of-the-art approach on most of the evaluation metrics. In terms of AUC, \toolname{} outperforms the best baseline by 4.9\% and 13.7\% in Java and Python, respectively.
    In terms of effort-aware metrics, i.e., CostEffort and $P_{opt}$, \toolname{} improves the best baseline up to 60\% and 51.4\%, respectively. 
    \item{We conduct two ablation studies and find that the designs of multi-level granularities and effort-aware adjustment are effective. Specifically, compared to single-level granularity, combining multiple granularities increases the performance up to 4.9\%, 8.5\% and 17.9\% in terms of AUC, CostEffort, and $P_{opt}$, respectively. Meanwhile, effort-aware adjustment boosts the performance of \toolname{} up to 21\% and 22\% in terms of CostEffort and $P_{opt}$, respectively.} 
\end{itemize}

\vspace{2mm}

\noindent \textbf{Organization.} The rest of the paper is organized as follows. 
Section \ref{sec:motivating} presents a motivating example that demonstrates the benefit of considering different levels of granularity for vulnerability-fixing commit detection.
Section \ref{sec:background} introduces background of the target problems and the used techniques.
Section \ref{sec:approach} describes the overview and main components of \toolname{}.
Section \ref{sec:evaluation} compares \toolname{} against other baselines for the target task.
Section \ref{sec:threats} mentions the threats to validity.
Section \ref{sec:related} introduces the related studies.
Finally, Section \ref{sec:conclusion} presents our conclusions and future directions.

\vspace{2mm}

\noindent \textbf{Data Availability.} To support the open science initiative, we published implementation and datasets of \toolname{} at

\begin{center}
    \url{https://github.com/soarsmu/midas}
\end{center}

\section{Motivating Example}
\label{sec:motivating}

\begin{figure*}
    \centering
    \includegraphics[width=1\textwidth]{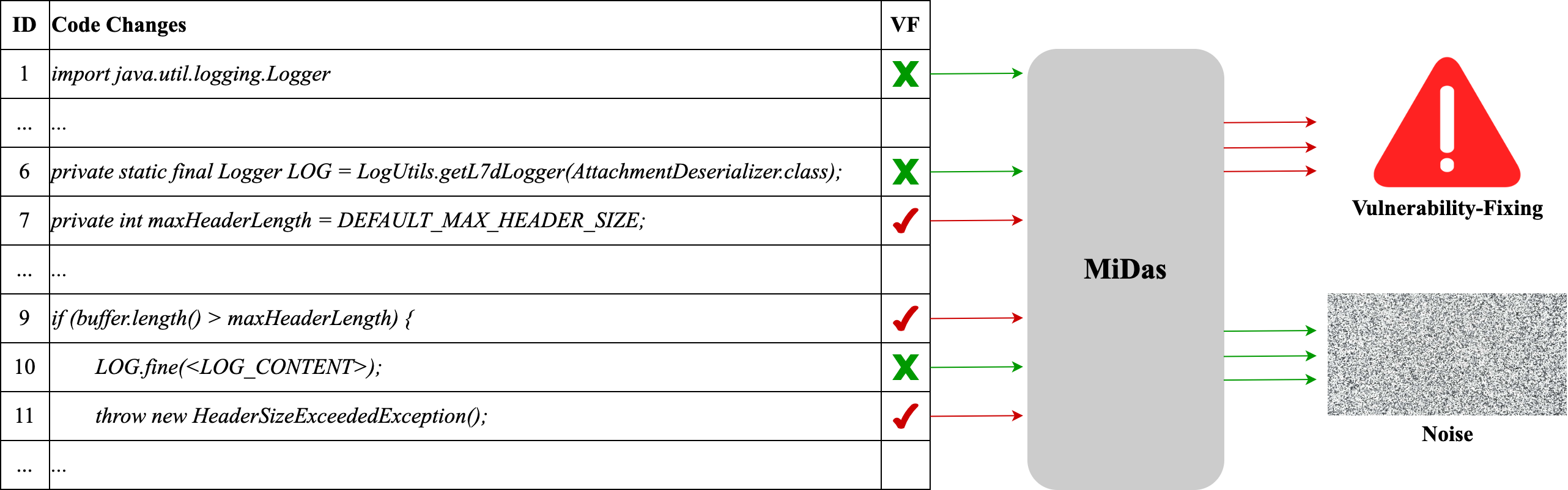}
    \caption{A sample commit fix at line-level granularity which is for fixing CVE-2017-12624 ("VF" denotes if the code change is related to implementing the fix)}
    \label{fig:line-level-motivation}
\end{figure*}

\begin{figure}[ht]
    \centering
    \includegraphics[width=0.5\textwidth]{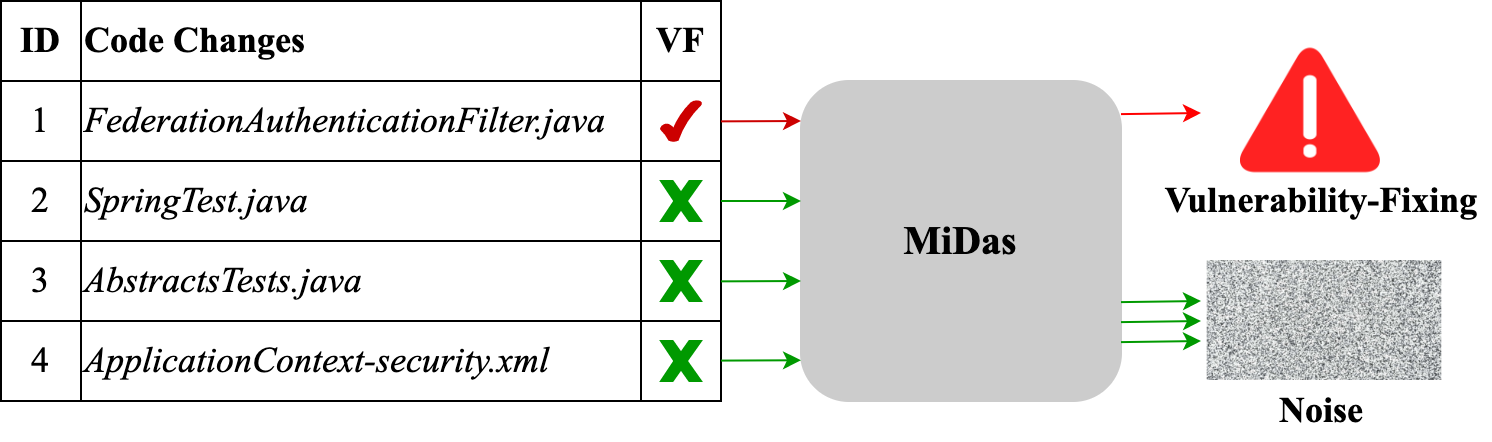}
    \caption{A sample commit fix at file-level granularity which is for fixing CVE-2017-12631 (``VF'' denotes if the code change is related to implementing the fix)}
    \label{fig:file-level-motivation}
\end{figure}

In this section, we present several motivating examples of vulnerability-fixing commits in the real applications to demonstrate the benefits of considering a commit as multi-level granularity structure data to achieve an effective classification.

Figure~\ref{fig:line-level-motivation}
presents a real-world commit made in Apache CXF~\footnote{\url{https://github.com/apache/cxf/commit/8bd915bfd7735c248ad660059c6b6ad26cdbcdf6}} which is to fix the vulnerability CVE-2017-12624~\footnote{\url{https://nvd.nist.gov/vuln/detail/CVE-2017-12624}}, a Denial of Service (DoS) vulnerability. 
The root cause of the vulnerability is directly from the improper logic handling related to the constant DEFAULT\_MAX\_HEADER\_SIZE in the source code.
As we see, the code changes in this commit spread across multiple files, hunks, and lines. 
However, we find that the key to determining whether the commit fixes the vulnerability or not is paying attention to the code changes at line-level, which serves to fix the root cause.
We observe that the remaining code changes are for other purposes, like logging and testing.
For all the aforementioned reasons, we believe that either using commit-level, file-level, or hunk-level granularity is not suitable to represent code changes because applying embedding models at these levels would possibly return noisy features. Indeed, the state-of-the-art model\cite{zhoufinding}, which represents code changes at file-level granularity, failed to classify this commit as a vulnerability-fixing commit.

Figure~\ref{fig:file-level-motivation} shows another example commit in a real application~\footnote{\url{https://github.com/apache/cxf-fediz/commit/48dd9b68d67c6b729376c1ce8886f52a57df6c4}}. The commit is to fix the vulnerability CVE-2017-12631~\footnote{\url{https://nvd.nist.gov/vuln/detail/CVE-2017-12631}}, which is related to Cross Style Request Forgery (CSRF). The commit contains four file changes, where only one of them is dedicated to implementing the fix, and the remaining two files are for testing. 
In other words, the commit is \emph{tangled}, and this phenomenon has been proved to be common \cite{barnett2015helping}.
Prior works \cite{nguyen2022hermes, sabetta2018practical} process all the code changes within a commit without recognizing their source files. In such a way, the code changes in the test files considered as noises in this example will be mixed with the code changes for vulnerability fixing.
Thus, we find that considering the code changes of a commit at file-level granularity can help separate the code changes for vulnerability fixing from other purposes. As a result, it could further boost the performance for our target task, i.e., vulnerability-fixing commit classification.

From the above examples, they motivate us to consider features from multiple levels of granularity for the vulnerability-fixing commit classification.

\section{Background}
\label{sec:background}
\begin{figure}[t]
\includegraphics[width=\columnwidth]{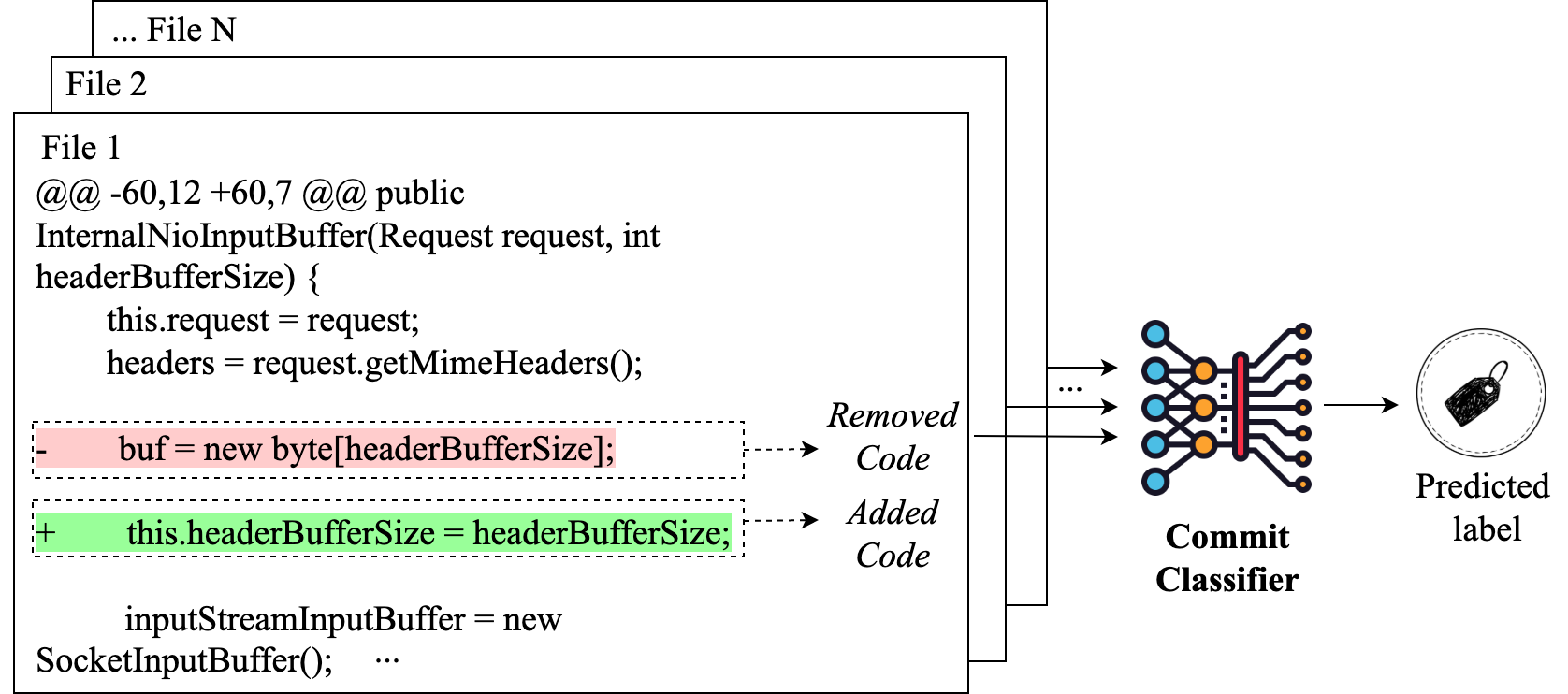}
\centering
\caption{Input/Output of Vulnerability-fixing Commit Classification}
\label{fig:vulguider_input_output}

\end{figure}

In this section, we first present the formal definition of the problem. And then, we introduce the essential background of the different types of neural networks leveraged in our approach.

\begin{figure*}[htbp]
\includegraphics[width=2\columnwidth]{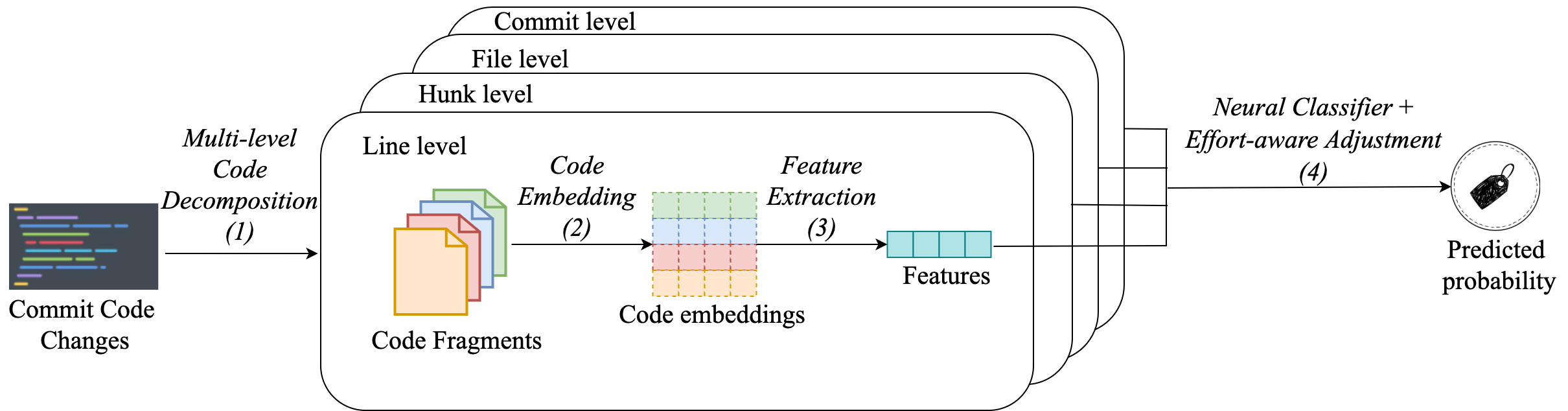}
\centering
\caption{Overview of \toolname{}}
\label{fig:overview}
\end{figure*}

\subsection{Vulnerability-fixing Commit Classification
}

Following many previous works, we formulate the vulnerability-fixing commit classification task as a binary classification problem. 
Formally, the input and output of the problem are described as follow (see Figure \ref{fig:vulguider_input_output}):

\vspace{1mm}

\noindent \textbf{Input: a commit.} In this task, we only consider the code changes of a commit as our input and ignore other information, e.g., commit message, by following the prior work \cite{zhoufinding}. 
The code changes may spread across multiple files, where code changes on the single file could consist of one or multiple hunks (i.e., groups of differing lines). Each hunk is in the form of a group of added and removed lines of code.

\vspace{1mm}

\noindent \textbf{Output: whether the commit is for vulnerability-fixing or not.}
Many existing approaches derive the output by producing a probability from 0 to 1 as the likelihood that the commit is for vulnerability-fixing. The higher the probability, the more likely the commit is a vulnerability-fixing commit.
\subsection{CodeBERT}
CodeBERT \cite{feng2020codebert} is a bimodal pre-trained model for programming language (PL) and natural language (NL) \cite{vaswani2017attention}.
It is trained on a large-scale dataset CodeSearchNet~\cite{husain2019codesearchnet} written in six programming languages, Python, Java, JavaScript, PHP, Ruby, and Go, respectively. The dataset consists of over 2.1M bimodal datapoints, which refers to pairs of NL-PL, and 6.1M unimodal datapoints, which refers to only PL.
% \bowen{I added following content, pls check.}

CodeBERT considers two tasks at the pre-training stage: masked language modeling (MLM) and Replaced Token Detection (RTD). Briefly, given an input sentence where some tokens are masked out, the MLM task predicts the original tokens for those masked tokens. 
For the RTD task, it aims to identify which tokens are replaced from the given input. 
The bimodal datapoints are used for both tasks, whereas the RTD task further uses unimodal datapoints to train the model. Hence, CodeBERT is able to handle both modalities of data. The model has been proven practical in various SE-related downstream tasks, such as natural language code search~\cite{codebert_base}, code document generation~\cite{codebert_base, yu2022bashexplainer}, program analysis~\cite{le2022autopruner, kazerounian2021simtyper} and program repair~\cite{mashhadi2021applying, xia2022less, le2023invalidator}.

\subsection{Deep Neural Networks}
\subsubsection{Convolutional Neural Network (CNN)}
CNN \cite{lecun1998gradient} is a type of neural network for extracting high-level features from input data. To achieve this, a CNN model first employs convolutional layers to generate the connectivity of local input features via kernels, which are $K \times K$ weight filters. Particularly, an input and its adjacent features are multiplied with a linear filter and then summed before being added a bias term and passed through an activation function such as ReLU~\cite{nair2010rectified} or Sigmoid~\cite{han1995influence}. In this way, convolutional layers can capture the local correlation of the inputs. Moreover, to empower convolutional layers, CNN uses a pooling mechanism, which partitions the output of convolutional layers into several non-overlapping regions and outputs the max, min, or average of each region. The mechanism enables CNN to reduce the feature dimensions as well as keep important features.
CNN has been proven its effectiveness in many SE tasks, like software question and answering posts representation~\cite{xu2021post2vec}, fault localzation \cite{li2021fault}, code generation \cite{sun2019grammar}, or just-in-time defect prediction \cite{hoang2019deepjit}

\subsubsection{Long Short-term Memory (LSTM)
}

LSTM \cite{hochreiter1997long} is a special kind of Recurrent Neural Networks (RNNs) capable of handling long-term dependencies in sequential data. 
A standard LSTM unit comprises a forget gate, an input gate, an output gate, and a memory cell. 
The forget gate decides information from memory that is forgotten, the input gate selects new information to update the memory, and the output gate controls the extent to the information in the memory to update the hidden state of the LSTM unit. 
In this way, LSTMs regulate the information that should be kept or discarded while traveling through the data sequence to avoid the problem of long-term dependencies.
In this paper, to enhance the learning capability of the model,  we employ an extension of LSTM, i.e., Bidirectional LSTM \cite{graves2005framewise}, which enables additional training via traversing the input data twice: left-to-right and right-to-left. 

\section{Approach}
\label{sec:approach}

Figure \ref{fig:overview} illustrates the overall architecture of our proposed approach for detecting vulnerability-fixing commits, namely \toolname{}.
\toolname{} takes a commit as its input, then outputs the probability indicating that a commit is for vulnerability-fixing or not.
More specifically, \toolname{} consists of five steps:
\begin{itemize}
    \item \textbf{Multi-level code decomposition} extracts information from a commit at different levels of granularity, i.g, lines, hunks, files or a whole commit,
    \item \textbf{Code Embedding}
    encodes the extracted information at different levels of granularities into numerical vectors by using a pre-trained model as inputs to the deep learning models in the feature extraction layers.
    \item \textbf{Feature Extraction}
    extracts features of commit codes at each level of granularity. The features are then concatenated to form the final representation of the input commit.
    \item \textbf{Neural Classifier} learns the mapping from the final representation of the input commit to the corresponding output vector in the training stage and then infers the likelihood that the commit is for vulnerability fixing. 
    \item \textbf{Effort-aware Adjustment} adjusts output probability of neural classifier to guarantee our system performance with limited human efforts.
    \end{itemize}
    
In the rest of the section, we introduce each step with more details. 

\subsection{Multi-level Code Decomposition} \label{sec:decomposition}
By design, a commit is in the form of code changes applied on a set of files. Each hunk shows one area where the files differ and it is in the form of a sequence of code changes applied on lines of code (LOC).
Considering the structure of commits, our approach extracts information from a commit at different levels of granularities in this step.
To achieve this, it decomposes a commit into code fragments corresponding to four levels of granularity based on the natural organization of a commit, i.e., line, hunk, file, and commit.
For example, at line-level granularity, we split code changes into lines then treat each input commit as a sequence of LOC.
As a result of this step, we obtain representations of the input commit at four levels of granularity as follows:
\begin{itemize}
    \item \textbf{Commit-level}: A input commit is considered as a single code fragment by sequentially concatenating code changes of the whole commit.
    
    \item \textbf{File-level}: A input commit $C$ is considered as a set of code fragment, $C=\{f_{1}, f_{2}, ..., f_{F}\}$ where,
    \begin{itemize}
        \item $F$ is the number of files in the input commit
        \item $f_{i}$ is a code fragment created by sequentially concatenating code changes of the $i^{th}$ file in the input commit
    \end{itemize}
    
    \item \textbf{Hunk-level}: A input commit $C$ is considered as a set of code fragment, $C =\{h_{1}, h_{2}, ..., h_{H}\}$ where,
        \begin{itemize}
            \item $H$ is the number of hunks in the input commit
            \item $h_{i}$ is a code fragment created by sequentially concatenating code changes of the $i^{th}$ hunk in the input commit
        \end{itemize}

    \item \textbf{Line-level}: A input commit $C$ is considered as a set of code fragment, $C=\{ l_{1}, l_{2}, ..., l_{F}\}$ where,
    \begin{itemize}
        \item $F$ is the number of files in the input commit
        \item $l_{i}$ is a code fragment of the $i^{th}$ line in the input commit
    \end{itemize}
\end{itemize}

\begin{figure}[htbp]
\includegraphics[scale=0.3]{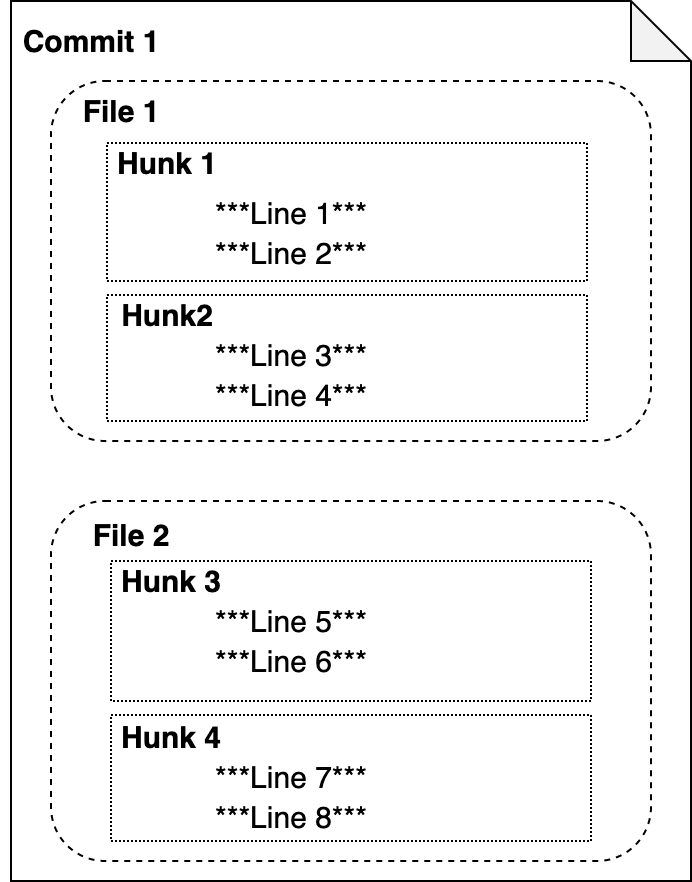}
\centering
\caption{An example for extracted code fragments for different granularity after multi-level code decomposition}
\label{fig:granularity_example}
\end{figure}

Figure \ref{fig:granularity_example} shows the structure of a commit. \textit{Commit 1}  involved changes in two files \textit{File 1}  and \textit{File 2}. Following that, \textit{Hunk 1} and \textit{Hunk 2} in \textit{File 1}, and \textit{Hunk 3}  and \textit{Hunk 4}  in \textit{File 2 } were modified, respectively. In every hunk, each of them contains 2 modified lines, from \textit{Line 1} to \textit{Line 8}. As the result of multi-level code decomposition, we obtain code fragments belong to different granularity as following:

\begin{itemize}
    \item \textbf{Commit-level}: $\{Commit\ 1\}$ 
    
    \item \textbf{File-level}: $\{File\ 1, File\ 2\}$ 
    
    \item \textbf{Hunk-level}: $\{Hunk\ 1, Hunk\ 2, Hunk\ 3, Hunk\ 4\}$   
    
    \item \textbf{Line-level}: $\{Line\ 1, Line\ 2, Line\ 3, Line\ 4, \\ Line\ 5, Line\ 6, Line\ 7, Line\ 8\}$   
\end{itemize}

\subsection{Code Embedding} \label{sec:code_embedding}
In this step, \toolname{} automatically represents code fragments as high-dimensional vectors. However, it faces a challenge in identifying vulnerability-fixing commits, which involves learning code representations automatically from a relatively small-scale dataset comprising less than 1,000 vulnerability-fixing commits. To overcome this challenge, \toolname{} leverages CodeBERT\cite{feng2020codebert}, which was pre-trained on a large-scale dataset and has shown good performance when fine-tuned with small datasets \cite{yang2022natural,mashhadi2021applying,xia2022less}. Specifically, \toolname{} first fine-tunes CodeBERT at each granularity level to capture the specific characteristics of code changes at each level, and then uses the fine-tuned models as code embedding models for representing code fragments.

By default, CodeBERT takes two segments as its input: one is for the data in natural language (NL), the other is in program language (PL). And its input is in the form of:

\begin{equation}
    [CLS] \langle \text{\textit{NL}} \rangle  [SEP] \langle \text{\textit{PL}} \rangle [EOS]
\end{equation}

where [\textit{CLS}], [\textit{SEP}], and [\textit{EOS}] are regarded as the special tokens in CodeBERT. Specifically, the [\textit{CLS}] token defines the start of a CodeBERT sequence, followed up by natural language text. The [\textit{SEP}] token is used to separate natural language text and program language source code. The [\textit{EOS}] token is put at the end of a CodeBERT sequence. For BERT-based models, e.g., CodeBERT, the network learns to generate meaningful embedding at the position of the [CLS] token during the training.

Recall that CodeBERT is pre-trained for two different modalities of data, which are \textit{bimodal data} (i.e., pairs of natural language and source code) and \textit{unimodal data} (i.e., source code). Hence, in our cases, we observe that code changes in an input commit, particularly, added code and removed code, can be considered in two different ways, considering the presence of source code context; we name them  \textit{context-dependent} and \textit{context-free} representation.

\vspace{2mm}
\noindent \textbf{Context-dependent representation.} In this representation, we consider \textit{removed code} and \textit{added code} within a code fragment as a pair of data.

This method aims to learn a joint representation of both the code added and removed in a commit. This representation contextualizes the added code with the removed code and vice versa. 

More formally, a code fragment will be represented in input format of CodeBERT as follows:
\begin{equation}
    [CLS] \langle \text{\textit{rem-code}} \rangle [SEP] \langle \text{\textit{add-code}} \rangle [EOS] 
\end{equation}
Then, we forward this representation to CodeBERT model and we take the output at [\textit{CLS}] token as \textit{initial embedding} of the code fragment.

\vspace{2mm}

\noindent \textbf{Context-free representation.} In this representation, we consider \textit{removed code} and \textit{added code} within a code fragment as two different unimodal datapoints. This representation treats removed code and added code separately without considering their counterparts. 

More formally, a code fragment will be represented in input format of CodeBERT as follows:
\begin{equation}
    [CLS] \langle \text{\textit{empty}} \rangle [SEP] \langle \text{\textit{add-code}} \rangle [EOS] 
\end{equation}
\begin{equation}
    [CLS] \langle \text{\textit{empty}} \rangle [SEP] \langle \text{\textit{rem-code}} \rangle [EOS] 
\end{equation}
Then, we forward these representations to CodeBERT model. In this case, we obtain two initial embeddings, one of \textit{added code} and one of \textit{removed code}, for the code fragment. Note that, CodeBERT can only take maximum 512 tokens. Hence, in case the input exceeds the limit, we truncate it by only consider the first 512 tokens. 

By combining four levels of granularity (as discussed in Section \ref{sec:decomposition}) and two different modalities of code fragments, we obtain seven settings of commit embedding as illustrated in Table \ref{tab:ref_base_settings}. Note that we leave the combination of line-level granularity and context-dependent representation for future work due to the fact that the combination requires an alignment between lines in \textit{removed code} and \textit{added code}, which are not available in the context of code changes \footnote{The existing tool of code alignment (a.k.a differencing) such as GumTree \cite{falleri2014fine}, however, the accuracy of such tool is not perfect\cite{fan2021differential} }. 

% by applying 4 levels of granularities and 2 ways of code fragment representation by CodeBERT, including 2 for commit-level, 2 for file-level, 2 for hunk-level, and 1 for line-level (using the second method). The detailed settings of 7 variants are illustrated in table \ref{tab:ref_base_settings}

\begin{table}[htbp]
\caption{Commit embedding settings}
\label{tab:ref_base_settings}
\begin{center}
\begin{tabular}{ccc}
\hline
% \cline{2-4} 
\textbf{Granularity} & \textbf{Representation}  & \textbf{Feature Extractor} \\
\hline
Commit & Context-dependent & FCN \\
File & Context-dependent & FCN \\
Hunk & Context-dependent & CNN \\
Commit & Context-free  & FCN \\
File & Context-free  & FCN \\
Hunk & Context-free  & CNN \\
Line & Context-free & LSTM \\
\hline

\end{tabular}\\ \vspace{2mm}

\end{center}
\end{table}

\subsection{Feature Extraction}

We observed that the characteristics of four levels of granularity differ. Thus, to effectively extract features from each of them, we utilize different models accordingly.
Overall, the feature extraction for each level of granularity follows a common structure consisting of a feature extractor followed by a feature fusion layer.
Note that each base model has one feature extractor and one feature fusion layer, where the feature extractor's design is customized for each granularity, and the feature fusion layer is shared by different granularity. We present these steps in detail below.

\subsubsection{Feature Extractor}
\label{sec:feature_extractor}

We leverage four deep learning models as feature extractors for different levels of granularity. 
For a commit, each feature extractor takes embedding vectors corresponding to the specific granularity as input and returns a feature vector as output.
We present the detailed architecture of these models as follows. 

\vspace{2mm}

\noindent \textbf{Line-level}:
Since lines between code changes are read sequentially, we leverage a Recurrent Neural Network,  a standard model for processing sequential data, to extract features at the line-level granularity. Particularly, we treat code changes as a sequence of lines, where each line is represented by an embedding vector as described in Section \ref{sec:code_embedding}, denoted as $[l_{1}, l_{2}, ..., l_{T_{lines}}]$. And then, we employ Bi-directional LSTM (BiLSTM) model as our feature extractor.
In our case, the bi-directional LSTM uses a forward LSTM that reads the commit from $l_{1}$ to $l_{T_{line}}$ and a backward LSTM that reads the commit from $l_{T_{line}}$ to $l_{1}$. We obtain final output of LSTM as the features of the commit.
\begin{equation}
    f_{line} = BiLSTM([l_{1}, l_{2}, ..., l_{T_{line}}])
\end{equation}

\vspace{2mm}

\noindent \textbf{Hunk-level}: 
Different from lines, the hunks within a commit do not carry sequential relationship.
However, there are still dependencies between hunks that are close, for example, hunks that are in the same file. The dependencies can be shared variables, constants or function calls.
Hence, we use a Convolutional Neural Network, which has demonstrated its ability to capture local dependencies in many tasks, e.g., sentences modeling \cite{kalchbrenner2014convolutional} or face recognition \cite{lawrence1997face}.
Specifically, given a set of embedding vectors of hunk-level code fragments decomposed from a commit, denoted as $[h_{1}, h_{2}, ..., h_{H}]$, we first employ convolution layers aggregate information from neighboring hunks. More formally, the features of $i-th$ hunk-level embedding vector is represented by aggregating information from neighboring embedding vectors as, 
\begin{equation}
    h'_{i} = Conv([h'_{i-(w-1)/2}, ..., h'_{i+(w-1)/2}])
\end{equation}
where $w$ is a kernel size of the convolution layer $Conv$.
Then, we employ a max-pooling layer to extract the most important features from the input embedding vectors and obtain the final output as the features,
\begin{equation}
    f_{hunk} = MaxPool([h'_{1}, h'_{2}, ..., h'_{H}])
\end{equation}

\vspace{2mm}

\noindent \textbf{File-level}: At the file level, we aim to capture high-level relationships between all code in the commit. Therefore, we use a Fully Connected Neural Network (FCN) to capture the relationships of all files in a commit simultaneously. 
Specifically, give a set of embedding vectors of file-level code fragments decomposed from a commit as $[f_{1}, f_{2}, ..., f_{F}]$, we first represent the commit by concatenating features of all vectors from $f_{1}$ to $f_{F}$. As a result, we obtain a $n \times F$ dimensional vector as the representation of the commit, n is the vector dimension of each file (i.e., output size of codeBERT). Note that a fully connected layer often requires a fixed size of input features. Hence, to deal with the problem, we set $F$ as a predefined parameter. For each commit, if its number of files is smaller than $F$, we add some blank files so that all commits have the same number of files. Otherwise, we truncate it to only its first 
$F$ files. After obtaining a fixed size input vector, we employ a fully connected layer to obtain the output features, as follows:
\begin{equation}
    f_{file} = FCN(f_{1} \oplus f_{2} \oplus \cdots \oplus f_{F})
\end{equation}
where $\oplus$ is the concatenation operator, and FCN is a fully connected layer with an input size of $n \times F$.
\vspace{2mm}

\noindent \textbf{Commit-level}: Similar to file-level feature extraction, we also use a fully connected layer. Specifically, given $x$ is the commit-level embedding vector of a given commit produced by CodeBERT. We employ a fully connected layer to obtain the output features, as follows:
\begin{equation}
    f_{commit} = FCN(x)
\end{equation}
where FCN is a fully connected layer with input size and output size of $n$ with n is the size of $x$, i.e., output size of codeBERT.

\vspace{2mm}

\subsubsection{Feature Fusion} \label{sec:extractor}

Based on the extracted feature vectors, we further construct a set of fully-connected layers as our feature fusion.
Note that, due to the different types of code embedding discussed in Section \ref{sec:decomposition}, we have two different feature fusion, i.e., bimodal and unimodal fusion corresponding two representation (i.e., \textit{context-dependant} and \textit{context-free} representation) methods as follows:
\begin{itemize}
    \item \textit{Bimodal fusion}: As mentioned in the previous section, we only obtain one feature vector for context-dependant representation. Thus, we directly feed it to a linear layer to fuse the features.
    
    \item \textit{Unimodal fusion}: In this case, we have two feature vectors, one for \textit{added code} and one for \textit{removed code}. Hence, we first concatenate them into one vector then feed the vector into a linear layer to fuse the features.
\end{itemize}

\subsection{Classifier and Effort-aware Adjustment}
\label{classifier_effort_aware}

\subsubsection{Neural Classifier}
Given extracted features of a commit from our extractors (as discussed in Section \ref{sec:extractor}), we use a neural network classifier to indicate whether the commit is for vulnerability-fixing or not. 
To achieve that, we first concatenate features of a commit, which is extracted at multiple granularities, then forward it into two fully connected layers to estimate a probability that the given commit is for fixing a vulnerability.

\subsubsection{Effort-aware Adjustment} \label{sec:adjustment} 

To increase the number of detected vulnerability-fixing commit under a limited inspection cost,  i.e., the inspected line of codes (LOC), we propose an \textit{effort-aware adjustment} as a post-processing step.
The adjustment aims to adjust the output of our vulnerability-fixing classifier based on the length of commit to prioritize the shorter vulnerability-fixing commits over the longer ones.
Specifically, our effort-aware adjustment is defined as follows:

\begin{equation}
\label{eq:1}
    \mathcal{P}(c) = prob_{c}\times f(loc_c)
\end{equation}

Where P(c) is the adjustment applied to the probability predicted by the neural classifier, denoted as $prob_{c}$, for a given commit \textit{c}. We want P(c) to be proportional to the number of LOC of \textit{c}, $loc_{c}$. The greater the number of LOC, the greater the adjustment. Therefore, we denote $f(loc\_c)$ as a function of $loc_c$ that would satisfy this property. Nevertheless, $f(loc_{c})$ should be carefully designed so that the adjustment does not dominate the probability predicted by the neural classifier. Hence, we choose the logarithm function as our $f$ function. More formally,

\begin{equation}
\label{eq:2}
    f(loc_c) = log_{a}(loc_{c})
\end{equation}

In Equation \ref{eq:2}, \textit{a} is the maximum number of LOCs of the vulnerability-fixing commits in the training dataset. As $loc_{c}$ is greater or equal to 1 and less than \textit{a}, $log_{a}(loc_{c})$ is bounded from 0 to 1 for any commit in the training dataset. As a result, we have modified Equation \ref{eq:1} as follows: 

\begin{equation}
\label{eq:3}
    \mathcal{P}(c) = prob_{c}\times log_{a}(loc_{c})
\end{equation}

Based on proposed effort-aware adjustments, we adjust the output probability of neural classier to obtain the final score of each commit as follows:

\begin{equation}
\label{eq:4}
    \mathcal{S}(c) = prob_{c} - \mathcal{P}(c)
\end{equation}

Where \textit{c} is a given commit, $prob_{c}$ is the output probability of the neural classifier, and P(c) is the calculated value of effort-aware adjustment for \textit{c}. However, in the real world, there may be vulnerability-fixing commits with lengths greater than a. It would lead to a negative \textit{S(c)} in Equation \ref{eq:4}. As we favor shorter commits for inspection, these large commits will be poorly ranked; thus, we ignore these outliers. To preserve the correctness of our evaluation, we limit \textit{S(c)} to 0. Hence, Equation \ref{eq:4} can be written as follows:

\begin{equation}
\label{eq:5}
    \mathcal{S}(c) = \operatorname{max}(prob_{c} - \mathcal{P}(c), 0)
\end{equation}

To summarize, our effort-aware adjustment function will modify the predicted probabilities of all commits in the test dataset. This modification affects probability-based evaluation metrics, including AUC, CostEffort, and $P_{opt}$, which we will discuss further in Section \ref{sec:evaluation}.

\subsection{Training}
In this section, we discuss about the process of training \toolname{}, including training strategy and optimization.

\begin{table*}[t]
\caption{Statistics of Zhou et al. \cite{zhoufinding} dataset}
\label{tab:ref_regex}
\begin{center}
\begin{tabular}{c|cccc|cccc|c}
\hline
% \cline{2-4} 
\multicolumn{10}{c}{\textbf{Training Set}} \\
\hline
\textbf{Lang} & \multicolumn{4}{c|}{\textbf{V.F.}} &
\multicolumn{4}{c|}{\textbf{N.V.F.}} & \#Projects \\
& \#Commit & \#File & \#Hunk & \#Line & \#Commit & \#File & \#Hunk & \#Line & \\
\hline
Java & 983 & 2,011 & 7,205 & 35,423 & 31,323 & 74,661 & 281,656 & 1,314,231 & 120 \\ 
Python & 522 & 747 & 2,124 & 8,769 & 20,362 & 27,737 & 75,618 & 294,982 & 84 \\
\hline
\multicolumn{10}{c}{\textbf{Validation Set}} \\
\hline
\textbf{Lang} & \multicolumn{4}{c|}{\textbf{V.F.}} &
\multicolumn{4}{c|}{\textbf{N.V.F.}} & \#Projects \\
& \#Commit & \#File & \#Hunk & \#Line & \#Commit & \#File & \#Hunk & \#Line & \\
\hline
Java & 191 & 224 & 798 & 3,801 & 6,921 & 8,296 & 31,106 & 147,286 & 119 \\ 
Python & 80 & 83 & 240 & 916 & 2,949 & 3,082 & 8,744 & 32,450 & 83 \\
\hline
\multicolumn{10}{c}{\textbf{Testing Set}} \\
\hline
\textbf{Lang} & \multicolumn{4}{c|}{\textbf{V.F.}} &
\multicolumn{4}{c|}{\textbf{N.V.F.}} & \#Projects \\
& \#Commit & \#File & \#Hunk & \#Line & \#Commit & \#File & \#Hunk & \#Line & \\
\hline
Java & 300 & 689 & 2,522 & 11,346 & 87,856 & 208,363 & 859,385 & 3,670,328 & 30 \\ 
Python & 195 & 254 & 613 & 2,384 & 55,638 & 72,752 & 205,763 & 784,006 & 22 \\
\hline
\end{tabular}\\ \vspace{2mm}
V.F.: Vulnerability-fixing Commits, N.V.F.: Non-vulnerability-fixing Commits.

\end{center}
\end{table*}

\subsubsection{Training Strategy}

As mention in Section \ref{sec:approach}, \toolname{} employs multiple feature extractors, corresponding to different commit embedding settings (refer to Table \ref{tab:ref_base_settings}). 
Technically, fully training \toolname{} is too expensive because it would require extensive resources of hardware and time. 
Therefore, we split the training process of \toolname{} into two phases, namely Base Model Training and Ensemble Training, respectively.
In Base Model Training, we independently train each base model which corresponds to each commit embedding setting.
Next, in Ensemble Training, we use a neural classifier to combine output features from these base models to initially obtain the predictions from \toolname{} .

\vspace{2mm}

\noindent \textbf{Base Model Training} The target of this phase is to train base models, in which each model consists of a CodeBERT and a feature extractor, to classify commits with respect to the corresponding embedding setting.
Ideally, we want to train each base model in one fold. 
However, because using CodeBERT is resource-expensive, one-fold training is only applicable for base models in which the number of code fragments for one commit is small, i.e., commit-level and file-level base models.
In other base models (i.e., line-level and hunk-level base models), we split this training phase into two steps.
The first step is to fine-tune CodeBERT to predict if a code fragment is for vulnerability-fixing or not.
As our dataset contains only the ground-truth label for the entire commit, to finetune CodeBERT, we heuristically consider that a code fragment is vulnerability-related if it belongs to a vulnerability-fixing commit. After finetuned, we freeze all CodeBERT's parameters and use embedding extracted by CodeBERT to train the corresponding feature extractor.  

\vspace{2mm} 

\noindent \textbf{Ensemble Training} In this phase, we freeze all parameters of base models, which are pre-trained in the previous phase, and only train the neural classifier.

\subsubsection{Optimization}
As \toolname{} is a vulnerability-fixing commits detector, which solves the problem belonging to binary classification, our training objective is to minimize the Cross-Entropy for the model on the entire training dataset. To update the weights of our neural networks, we use Adam optimizer \cite{kingma2014adam}, which is broadly used in many fields of deep learning. The learning rate is set to 1e-5 following CodeBERT \cite{feng2020codebert}. 

For base model training, CodeBERT of each base model is fine-tuned for one epoch. After that, each base model is trained on training set. The process stops training if the value of the Cross-Entropy loss on the validation set has not been updated in the last five epochs. All base models are trained for a maximum of 60 epochs. For ensemble training, the neural classifier is trained with a learning rate of 1e-5 and 20 epochs.

\subsection{Application}
In an industrial setting, vulnerability-fixing commits detected through machine learning undergo a manual assessment by human experts \cite{zhou2017automated, sabetta2018practical}. Our proposed approach \toolname{} supports the same setting, aiding security experts/researchers in monitoring commits. Given a set of commits as inputs, \toolname{} outputs a ranked list of possible vulnerability-fixing commits.
Previous studies have suggested that security experts can leverage commits that address potential vulnerabilities to enhance IT supply chain security within the industry \cite{zhou2017automated, zhoufinding, ponta2019manually, sabetta2018practical}. For instance, Zhou and Sharma's approach \cite{zhou2017automated} was utilized to identify vulnerability-fixing commits for developing Software Composition Analysis (SCA) database in Veracode. Similarly, Sabetta et al. \cite{ponta2019manually, sabetta2018practical} extended this work at SAP, creating the SCA database for their vulnerability assessment tool, Eclipse Steady\footnote{\url{https://eclipse.github.io/steady/}
}. Additionally, SAP developed Prospector\footnote{\url{https://github.com/SAP/project-kb/tree/commit-in-adv/prospector}}, which utilizes a vulnerability description in natural language as input to produce a ranked list of commits in decreasing order of relevance, thereby reducing the effort required to identify security fixes for known vulnerabilities in open-source software repositories. Zhou et al.~\cite{zhoufinding} further extended this research to develop VulFixMiner, a vulnerability-fixing commit identification model for Huawei, which is proven capable of detecting unreported vulnerability-fixing commits as confirmed by security experts. In this paper, we demonstrate that our solution, \toolname{}, significantly outperforms the state-of-the-art VulFixMiner across multiple programming languages.

\section{Evaluation}
\label{sec:evaluation}

Our experiments are driven by the following research questions (RQs):

\vspace{0.2cm}\noindent{\bf RQ1. How effective is \toolname{} compared to the baselines?} To answer this RQ, we compare \toolname{} with VulFixMiner~\cite{zhoufinding}, the current state-of-the-art approach, which is also designed for vulnerability-fixing commit classification. We also utilized LApredict ~\cite{zeng2021deep} and DeepJIT~\cite{hoang2019deepjit}, which are the state-of-the-art approaches for buggy commit detection (e.g., JIT defect prediction).
Furthermore, we investigate the technical differences between \toolname{} and the state-of-the-art baseline. We analyze the components of \toolname{} and compare the performance of different versions of \toolname{} with the state-of-the-art baseline.

\vspace{0.2cm}\noindent{\bf RQ2. How does the effort-aware objective function affect the performance of \toolname{}?} This RQ aims to investigate the contribution of our effort-aware objective function to \toolname{}. We answer the question by comparing the performance of \toolname{} in two versions, with and without the effort-aware objective function, respectively.

\vspace{0.2cm}\noindent{\bf RQ3. How do different levels of granularity affect the performance of \toolname{}?} The goal of this RQ is to investigate the influence of different levels of granularity on the performance of \toolname{}. We answer this RQ by continuously combining base models corresponding to each level of granularity and evaluating their performance on the considered evaluation metrics.

\vspace{0.2cm}\noindent{\bf RQ4. Can \toolname{} detect vulnerability-fixing commits that involve different types of changes?} To answer this question, we evaluate \toolname{} on commits containing 5 or more hunks. And then, we evaluate the performance of \toolname{} in comparison with the state-of-the-art baseline on the sub-datasets.

\subsection{Dataset}

To facilitate comparison, we evaluate \toolname{} on the dataset proposed by VulFixMiner \cite{zhoufinding} and follow exactly their dataset configuration. The dataset contains both vulnerability-fixing and non-vulnerability-fixing commits extracted from 150 Java projects and 106 Python projects.
The vulnerability-fixing commits were collected from two sources. The first source is a manually curated Java vulnerability-fixing commit dataset, namely the SAP dataset \cite{ponta2019manually}. The SAP dataset contains 1,055 vulnerability-fixing commits, spanning 183 Java OSS projects. 
These projects were identified based on data analysis at SAP while operating their vulnerability assessment tool called Vulas. The corresponding vulnerability-fixing commits were then manually collected over a period of four years by monitoring the disclosure of security advisories, not only from NVD, but also from projects-specific web pages. The dataset is verified by SAP researchers based on several resources such as code changes, commit messages, and reference issues. 

The second source is all CVEs related to Java and Python disclosed by January 26, 2021. From the CVEs, Zhou et al.\cite{zhoufinding} collected 199 commits, 227 issues, 155 pull requests in Java, 288 commits, 244 issues, and 353 pull requests in Python. Then, the commits referenced in the pull requests and issues are extracted. Finally, all commits are merged into a single dataset after removing duplicate commits. For non-vulnerability-fixing commits, commits are sampled from the projects containing vulnerability-fixing commits up until February 26, 2021.

Until this point, the Java dataset contains 1,436 vulnerability-fixing commits and 839,682 non-vulnerability-fixing commits. Meanwhile, the Python dataset contains 885 vulnerability-fixing commits and 722,291 non-vulnerability-fixing commits. Afterward, Zhou et al.~\cite{zhoufinding} further filtered the dataset by removing large commits that are less likely to fix vulnerabilities. The removal resulted in 474,555 non-vulnerability-fixing commits, and 1,353 vulnerability-fixing commits from 150 projects for Java. For Python, the corresponding values are 357,696 non-vulnerability-fixing commits and 751 vulnerability-fixing commits from 106 projects. Finally, Zhou et al. \cite{zhoufinding} enhance the dataset by labeling more commits that are relevant to vulnerability fixes, more specifically, commits which message contains vulnerability-related keywords (i.e., “vuln”, “CVE”, and “NVD”). To ensure the pattern is well-designed, Zhou et al.~\cite{zhoufinding} randomly sampled a subset of extracted commits by it and manually verified them. As a result, in the Java dataset, they relabel 420 non-vulnerability-fixing commits across 123 projects. In the Python dataset, they relabel 501 non-vulnerability-fixing commits across 98 projects.

The dataset follows the standard manner that it is split into three parts without overlap of projects, training set, validation set, and testing set. Recall the dataset configuration, the dataset is split project-wise, using an 80\%/20\% split and consider the 20\% split as testing dataset. Then, the remaining 80\% is further split with the ratio 90\%/10\%, consider using 90\% for training dataset and 10\% for testing dataset. Note that the training and validation dataset are randomly under-sampled to reduce the imbalanced nature.
The details of the dataset distribution are shown in Table~\ref{tab:ref_regex}.

\subsection{Evaluation Metrics}
\label{sec:metrics}

To facilitate a fair comparison, we use the same evaluation metrics by following the prior work ~\cite{zhoufinding}, they are AUC and two effort-aware metrics (i.e., CostEffort@L and $P_{opt}$@L).

\textbf{AUC} (\underline{A}rea \underline{U}nder the \underline{C}urve): is the area under the Receiver Operating Characteristic (ROC) Curve \cite{hanley1982meaning}. It is a threshold-independent measure, which illustrates the discriminant ability of proposed techniques for binary classification problem \cite{lessmann2008benchmarking}. AUC represents the probability that a randomly chosen negative example (i.e., non-vulnerability-fixing commit) will be ranked higher than a randomly chosen positive example (i.e.,vulnerability-fixing commit). More formally, AUC score is calculated as follow:
\begin{equation}
    AUC = \dfrac{S_{0} - n_{0}(n_{0} + 1)/2}{n_{0}n_{1}}
\end{equation}
where $n_{0}$ and $n_{1}$ are the numbers of vulnerability-fixing and non-vulnerability-fixing commits, respectively, and $S_{0} = \Sigma{r_{i}}$, where $r_{i}$ is the rank of the $i^{th}$ vulnerability-fixing commit in the descending list of output probability produced by each model. 

\vspace{2mm}

\textbf{CostEffort@L}: The goal of a vulnerability-fixing commit detector is to rank vulnerability-fixing commits higher than the non-vulnerability fixing ones, so that, developers are capable of inspecting the code changes (i.e., the number of inspected lines of code) with a specific amount of effort. Given the commits, which are ordered by predicted probabilities obtained from the model, CostEffort@L counts the number of detected vulnerability-fixing commits, starting from commit with high to low predicted probabilities until the number of lines of code changes reaches L\% of total LOC. The value of CostEffort@L represents the effectiveness of the approach under the predefined inspecting cost. The higher value of CostEffort@L, the better effectiveness of the model. In~\cite{zhoufinding}, only CostEffort@5\% and CostEffort@20\% are considered. In this work, we also calculate CostEffort@10\% and CostEffort@15\% to investigate how the performance differs with the increase of inspecting cost.

%%%%%%%%%%%%%%%%%%%%%%

\begin{table*}[htbp]
\caption{Performance of \toolname{} and baseline models on Java and Python projects}
\label{tab:rq1_all}
\begin{center}
\begin{tabular}{l|l|c|cccc|cccc}
\hline
\textbf{Lang} & \textbf{Model} & \textbf{AUC} & \multicolumn{4}{c|}{\textbf{CostEffort}} & \multicolumn{4}{c}{\textbf{$P_{opt}$}} \\
&  &  & 5\% & 10\% & 15\% & 20\% & 5\% & 10\% & 15\% & 20\% \\
 \hline
\multirow{4}{*}{\textbf{Java}} & \textbf{VulFixMiner} & 0.81 & 0.61 & 0.65 & 0.68 & 0.71 & \textbf{0.53} & 0.58 & 0.61 & 0.63 \\
& \textbf{DeepJIT} & 0.83 & 0.34 & 0.48 & 0.50 & 0.62 & 0.24 & 0.33 & 0.38  & 0.43 \\
& \textbf{LApredict} & 0.45 & 0.22 & 0.38 & 0.49 & 0.59 & 0.13 & 0.21 & 0.29 & 0.35\\
& \textbf{LOC-sensitive model} & 0.37 & 0.32 & 0.50 & 0.59 & 0.67 & 0.19 & 0.30 & 0.39 & 0.45\\
& \textbf{\toolname{}} & \textbf{0.85} & \textbf{0.64} & \textbf{0.77} & \textbf{0.87} & \textbf{0.91} & 0.50 & \textbf{0.60} & \textbf{0.67} & \textbf{0.73}\\
\hline
\multirow{4}{*}{\textbf{Python}} & \textbf{VulFixMiner} & 0.73 & 0.32 & 0.40 & 0.48 & 0.56 & 0.24 & 0.30 & 0.35 & 0.39 \\
& \textbf{DeepJIT} & 0.60 & 0.08 & 0.13 & 0.22 & 0.33 & 0.05 & 0.08 & 0.12  & 0.16 \\
& \textbf{LApredict} & 0.48 & 0.12 & 0.17 & 0.23 & 0.29 & 0.08 & 0.11 & 0.14 & 0.17\\
& \textbf{LOC-sensitive model} & 0.47 & 0.27 & 0.44 & 0.52 & 0.61 & 0.16 & 0.25 & 0.33 & 0.39\\
& \textbf{\toolname{}} & \textbf{0.83} & \textbf{0.47} & \textbf{0.64} & \textbf{0.74} & \textbf{0.81} & \textbf{0.33} & \textbf{0.45} & \textbf{0.53} & \textbf{0.59}\\
\hline
\end{tabular}

\end{center}
\end{table*}

\begin{table*}[htbp]
\caption{Performance of \toolname{} with and without effort-aware adjustment on Java and Python. \toolname{} and $\text{\toolname{}}_{NoAdj}$ denote \toolname{} with/without adjustment, respectively}
\label{tab:rq2_all}
\begin{center}
\begin{tabular}{l|l|c|cccc|cccc}
\hline
\textbf{Lang} & \textbf{Model} & \textbf{AUC} & \multicolumn{4}{c|}{\textbf{CostEffort}} & \multicolumn{4}{c}{\textbf{$P_{opt}$}} \\
&  &  & 5\% & 10\% & 15\% & 20\% & 5\% & 10\% & 15\% & 20\% \\
 \hline
\multirow{2}{*}{\textbf{Java}} &  \textbf{$\text{\textbf{\toolname{}}}_{NoAdj}$} & \textbf{0.86} & 0.57 & 0.68 & 0.73 & 0.82 & 0.44 & 0.53 & 0.6 & 0.64\\ 
& \textbf{\toolname{}} & 0.85 & \textbf{0.64} & \textbf{0.77} & \textbf{0.87} & \textbf{0.91} & \textbf{0.50} & \textbf{0.60} & \textbf{0.67} & \textbf{0.73}\\
\hline
\multirow{2}{*}{\textbf{Python}} &  \textbf{$\text{\textbf{\toolname{}}}_{NoAdj}$} & 0.83 & 0.39 & 0.57 & 0.65 & 0.70 & 0.27 & 0.39 & 0.46 & 0.52\\ 
& \textbf{\toolname{}} & 0.83 & \textbf{0.47} & \textbf{0.64} & \textbf{0.74} & \textbf{0.81} & \textbf{0.33} & \textbf{0.45} & \textbf{0.53} & \textbf{0.59}\\
\hline
\end{tabular}

\end{center}
\end{table*}

\begin{table}[t]
\caption{Number of inspected commits at different inspection cost of LOC for Java and Python projects. \toolname{} and $\text{\toolname{}}_{NoAdj}$ denote \toolname{} with/without adjustment, respectively}
\label{table:rq2_analyze_all}
\begin{center}
\begin{tabular}{l|l|c|c|c|c}
\hline
% \cline{2-4} 
\textbf{Lang} & \textbf{Model} & \textbf{5\%} & \textbf{10\%} & \textbf{15\%} & \textbf{20\%} \\
\hline
\multirow{3}{*}{\textbf{Java}} & $\text{\textbf{\toolname{}}}_{NoAdj}$ & 5,301 & 11,602 & 18476 & 25597 \\
& \textbf{\toolname{}} & 9,588 & 22,460 & 33,850 & 42,993 \\
\hline
& \textbf{Increment} & 81\% &	94\% &	83\% &	68\% \\
\hline
\multirow{3}{*}{\textbf{Python}} & $\text{\textbf{\toolname{}}}_{NoAdj}$ & 2,677 & 5,689 & 8,896 & 12,390 \\
& \textbf{\toolname{}} & 4,045 & 8,774 & 14,044 & 18,986 \\
\hline
& \textbf{Increment} & 51\% &	54\% & 58\% & 53\% \\
\hline
\end{tabular}

\end{center}
\end{table}

\begin{table*}[htbp]
\caption{Performance of \toolname{} on Java and Python when continuously adding granularities}
\label{tab:rq3_all}
\begin{center}
\begin{tabular}{l|l|c|cccc|cccc}
\hline
\textbf{Lang} & \textbf{Model} & \textbf{AUC} & \multicolumn{4}{c|}{\textbf{CostEffort}} & \multicolumn{4}{c}{\textbf{$P_{opt}$}} \\
&  &  & 5\% & 10\% & 15\% & 20\% & 5\% & 10\% & 15\% & 20\% \\
\hline
\multirow{4}{*}{\textbf{Java}} & \textbf{Commit} & 0.83	& 0.55 & 0.72 &	0.81 &	0.87 &	0.41 &	0.53 &	0.61 &	0.67 \\
& \textbf{File} & 0.81 & 0.55 & 0.68 & 0.75 & 0.85 & 0.42 & 0.51 & 0.58 & 0.64 \\
& \textbf{Hunk} & 0.84 & 0.59 & 0.73 & 0.81 & 0.89	& 0.46 & 0.57 & 0.64 & 0.69 \\
& \textbf{Line} & 0.81 & 0.60 & 0.71 & 0.82 & 0.88 & 0.46 & 0.56 & 0.63 & 0.68 \\
& \textbf{Line + Hunk} & 0.84 & 0.62 & 0.74 & 0.84 & 0.90 & 0.49 & 0.59 & 0.66 & 0.71\\
& \textbf{Line + Hunk + File} & 0.84 & 0.61 & 0.76 & 0.84 & 0.89 & \textbf{0.50} & \textbf{0.60} & \textbf{0.67} & 0.72\\
& \textbf{\toolname{} (Line + Hunk + File + Commit)} & \textbf{0.85} & \textbf{0.64} & \textbf{0.77} & \textbf{0.87} & \textbf{0.91} & \textbf{0.50} & \textbf{0.60} & \textbf{0.67} & \textbf{0.73}\\
\hline
\multirow{4}{*}{\textbf{Python}} & \textbf{Commit} & 0.82 & 0.39 & 0.57 & 0.68 & 0.77 & 0.27 & 0.37 & 0.46 & 0.53 \\
& \textbf{File} & 0.80 & 0.43 & 0.56 & 0.65 & 0.74 & 0.27 & 0.38 & 0.46 & 0.52 \\
& \textbf{Hunk} & 0.82 & 0.47 & 0.63 & 0.71 & 0.78 & 0.32 & 0.44 & 0.52 & 0.58 \\
& \textbf{Line} & 0.81 & 0.44 & 0.63 & 0.73 & 0.81 & 0.28 & 0.41 & 0.50 & 0.57 \\
& \textbf{Line + Hunk} & 0.82 & \textbf{0.48} & \textbf{0.65} & 0.70 &    0.77 & 0.30 & 0.44 & 0.52 & 0.58\\
& \textbf{Line + Hunk + File} & 0.81 & 0.42 & 0.62 & \textbf{0.74} & 0.79 & 0.28 & 0.41 & 0.50 & 0.57\\
& \textbf{\toolname{} (Line + Hunk + File + Commit)} & \textbf{0.83} & 0.47 & \textbf{0.64} & \textbf{0.74} & \textbf{0.81} & \textbf{0.33} & \textbf{0.45} & \textbf{0.53} & \textbf{0.59}\\
\hline
\end{tabular}

\end{center}
\end{table*}

\begin{figure}
    \centering
    \includegraphics[width=0.4\textwidth]{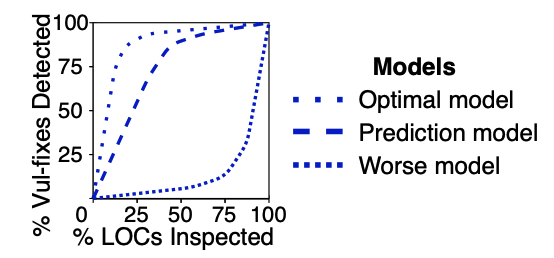}
    \caption{An example from Zhou et al.~\cite{zhoufinding} showing the relationship between the percentage of vulnerabilities fixes detected and the amount of inspection cost (i.e., $\%$ LOC) for different models}
    \label{fig:popt}
\end{figure}

\vspace{2mm}

\textbf{$P_{opt}$@L}:
$P_{opt}$ is a normalized version of the cost-aware performance metric introduced by Mende and Koschke \cite{mende2010effort}. Given an Alberg diagram \cite{ohlsson1996predicting} that shows the relationship between the number of vulnerability-fixing commits (on the y-axis) and the inspection cost (on the x-axis). $P_{opt}$@L is computed for a given inspection cost, L, which is the percentage of total lines of code (LOCs) inspected. $P_{opt}$ is an effort-aware performance metric used in studies on defect prediction \cite{kamei2012large, yu2019empirical, yang2016effort, huang2019revisiting}. $P_{opt}@L$ was also used in the previous study on detecting vulnerability fixing commits \cite{zhoufinding}

Assuming we wish to assess a prediction model M, which outputs a sorted list of commits. M is compared against the optimal model, O, and the worst model, W. Using the ground-truth labels, O and W order the commits as their output. O ranks ground-truth vulnerability-fixing commits higher than non-vulnerability-fixing commits, favoring  commits with fewer LOC. W ranks non-vulnerability fixing commits higher than vulnerability-fixing commits, favoring commits with a greater LOC. As such, the performance of the optimal model represents the upper bound of the performance of any prediction model, while the performance of the worst model represents the lower bound. $Curve_M$ , $Curve_O$, $Curve_W$ are the curves of the prediction model M, the optimal model O, and the worst model, W, respectively (see Figure \ref{fig:popt}). For any two models, A and B, Area($Curve_A$, $Curve_B$) is the corresponding area between the curves. The points on the curves for a given L correspond to the percentage of vulnerability-fixing commits detected with L\% of the total LOC inspected. For the prediction model M, $P_{opt}(M)$ is computed as: 

\begin{equation}
P_{opt}(m)=\dfrac{Area(Curve_M, Curve_W)}{Area(Curve_O, Curve_W)}
\end{equation}

A larger $P_{opt}$ value indicates that performance between the prediction model, M, is closer to the optimal model. In our experiments, we calculate $P_{opt}@L$ with four different values of L, which are 5, 10, 15, and 20.

\subsection{Baselines}\label{sec:baselines}

We compared \toolname{} with the following three baselines:

\vspace{2mm}

\noindent \textbf{VulFixMiner~\cite{zhoufinding}}:  VulFixMiner is the current state-of-the-art baseline in vulnerability-fixing commit identification.
It extracts commits at the file-level granularity and uses CodeBERT to represent code change of files. Embeddings of code changes of files are aggregated by an average function to form commit's embedding. Lastly, commit's embedding is used to train a neural classifier for prediction.
\vspace{2mm}

\noindent \textbf{DeepJIT~\cite{hoang2019deepjit}}: is a well-known deep learning approach for buggy commit identification (a.k.a defect prediction), which is relevant to our problem, i.e. vulnerability-fixing commit identification. DeepJIT takes inputs as a code change and commit message and uses deep learning models, i.e., Convolutional Neural Network, to predict whether a commit is defective or not. As our problem settings only involve code changes, we only use code change component of DeepJIT in our experiments for a fair comparison.

\vspace{2mm}

Other than the deep learning approaches, we compare \toolname{} with three simpler baselines. Sometimes, a simple model can outperform complex ones (e.g., deep learning neural networks) \cite{zeng2021deep, qi2014strength}. Hence, we add the two following baselines to our evaluation:

\vspace{2mm}

\noindent \textbf{LApredict}\cite{zeng2021deep}:LApredict is an approach using logistic regression with only one feature - the number of added LOCs. We selected LApredict as a simple baseline as it was shown to outperform a more complex approach \cite{hoang2020cc2vec} in identifying defective program changes. We compare \toolname{} with LApredict for two reasons. Firstly, LApredict is also proposed to address binary classification tasks with imbalanced data. Secondly, defects and vulnerabilities may potentially carry similar characteristics. Therefore, we want to know if LApredict can be generalized for our problem.

\vspace{2mm}

\noindent \textbf{LOC-sensitive model}: 
As introduced in Section~\ref{sec:metrics}, we consider CostEffort and $P_{opt}$ as our evaluation metrics. 
These two metrics assess the ability to detect vulnerability-fixing commits of the model based on the certain number of inspected LOCs.
Since under the same number of inspected LOCs, different models may inspect different numbers of commits, we are interested in investigating if a naive model that maximizes the number of inspected commits could yield a good result. The intuition is that under a fixed inspection cost, the more commits that are inspected, the more vulnerability-fixing commits are detected. 
To do that, this naive model simply ranks commits based on the number of LOC of code changes in ascending order. Under a fixed threshold of the total number of LOC, commits with the lower number of LOCs are inspected until the threshold is met.
In other words, the LOC-sensitive model assigns higher ranks for short commits than the long ones. 
As the other two baselines do not consider the amount of effort required, we use the LOC-sensitive model as a simple baseline that accounts for the amount of effort to make its prediction.

\subsection{Experiment Results}

\noindent\textbf{{RQ1. How effective is \toolname{} compared to the baselines?}}

To answer this question, we evaluate \toolname{} and baseline models on two datasets, in terms of AUC, CostEffort@k, $P_{opt}$@k (k equals 5, 10, 15, 20). Table \ref{tab:rq1_all} presents the performance results on Java and Python, respectively. Overall, \toolname{} outperforms all the baselines on all the evaluation metrics with one exception that VulFixMiner achieves the best performance on $P_{opt}$@5\%.

On the Java dataset, in terms of AUC, \toolname{} outperforms the best baseline, i.e., DeepJIT, by 2.4\% ((0.85-0.83)/0.83). Note that, except for this metric on Java, VulFixMiner is the best baseline on every metric on both Java and Python. In terms of CostEffort, the improvement achieved by MiDas over the best baseline, i.e, VulFixMiner, varies from 4.9\% to 28.2\% when the percentage of total LOC increases from 5\% to 20\% . Especially, with 20\% of LOC, \toolname{} can identify more than 90\% of the vulnerability fixes. On Popt, \toolname{} performs worse than VulFixMiner on Popt@5, but better by a large margin (i.e., 15.9\%) on Popt@20.
% \bowen{grammar issue, LOC cannot be a percentage}
. Especially, with 20\% of LOC, \toolname{} can identify more than 90\% of the vulnerability fixes. On $P_{opt}$, \toolname{} performs worse than VulFixMiner on $P_{opt}$@5, but better by a large margin (i.e., 15.9\%) on $P_{opt}$@20.

On the Python dataset, we find that the best performer among all the baselines is also VulFixMiner. Our model, \toolname{} outperforms VulFixMiner on all the metrics. In terms of AUC, \toolname{} leads an improvement by 13.7\% ((0.83-0.73)/0.73). For the effort-related metrics, \toolname{} outperforms VulFixMiner by a large margin varies from 45\% to 60\% and from 37.5\% to 51.4\% on CostEffort and $P_{opt}$, respectively.

Besides, LApredict and LOC-based-sorting-model perform poorly on all the metrics. It suggests that these approaches are ineffective for the vulnerability-fixing commits detection problem.
Since LApredict \cite{zeng2021deep} only considers one feature, namely the number of LOCs, we further determine if there is a correlation between the number of LOCs and whether a commit is intended to fix a vulnerability. To do so, we follow the approach taken in prior research \cite{kochhar2015code} and calculate the square of the point biserial correlation coefficient \cite{brown1988understanding}, denoted as \textit{spb} . The point biserial correlation coefficient is used to measure the correlation between two variables when one of them is dichotomous, taking values of either 0 or 1. To interpret the strength of the correlation, we use the interpretation given in existing studies \cite{kochhar2015code, pett2015nonparametric}, where $\textit{spb} \geq 0.81$ means a very strong correlation, $ 0.49 \leq \textit{spb} < 0.81 $ indicates a strong correlation, $ 0.25 \leq \textit{spb} < 0.49 $ indicates a moderate correlation, $ 0.09 \leq \textit{spb} < 0.25 $ indicates a weak correlation, and $ 0.00 < \textit{spb} < 0.09 $ indicates very weak correlation. The calculated \textit{spb} value is 0.00289, with a statistically significant p-value of less than 0.1, indicating a very weak correlation between the number of lines of code and whether a commit is a vulnerability fix.

From all the aforementioned results, we empirically illustrate that \toolname{}  has higher discriminative power in identifying vulnerability-fixing commits and is able to identify more vulnerability-fixing commits under the same inspection cost compared to all other baselines.

\vspace{2mm}

\noindent\textbf{{RQ2. How does the effort-aware adjustment affect the performance of \toolname{}?}}

To answer this RQ, we compared two versions of \toolname{}, with and without effort-aware objective function, respectively. The experimental results for Java and Python projects are mentioned in Table \ref{tab:rq2_all}, 
in which, \textbf{\toolname{}} denotes the performance of \toolname{} with the effort-aware adjustment, and $\text{\textbf{\toolname{}}}_{NoAdj}$ denotes the performance of \toolname{} without the effort-aware adjustment. Overall, although applying our effort-aware adjustment keeps AUC either remaining the same or decreases insignificantly (by 0.01), it improves the two effort-related metrics by a big margin.
Specifically, on the Java dataset, the improvement in CostEffort and $P_{opt}$ ranges from 11\% to 19.1\% and from 11.7\% to 14\%, respectively. The corresponding improvements for the Python dataset are from 12.3\% to 21\% and from 13.5\% to 22\%. The reason behind the improvement is that our effort-aware adjustment can boost the number of inspected commits without the loss of discriminative capability of \toolname{}, which is reflected by the stability of AUC. Indeed, as shown in Table \ref{table:rq2_analyze_all}, the effort-aware adjustment increases the number of inspected commits at least 68\% and 51\% for Java and Python projects, respectively.

Comparing the results of  $\toolname{}_{NoAdj}$ in Table \ref{tab:rq2_all} with the results of VulFixMiner in Table \ref{tab:rq1_all},
$\toolname{}_{NoAdj}$ outperforms the state-of-the-art baseline in terms of AUC, by 6.1\% ((0.86-0.81)/0.81) on Java and 13.7\% ((0.83-0.73)/0.73) on Python.
This improvement comes from the difference in the neural network design of \toolname{} and VulFixMiner. Specifically, VulFixMiner considers only file-level granularity. Meanwhile, \toolname{} considers multiple granularities including commit-level, file-level, hunk-level, and line-level granularity, as described in Section \ref{sec:approach}. While, $\toolname{}_{NoAdj}$ underperforms on some thresholds of CostEffort@L and $P_{opt}$, that are CostEffort@5\%, $P_{opt}$@5\%, $P_{opt}$@10\% on Java, by 6.6\% ((0.61-0.57)/0.61), 17\% ((0.53-0.44)/0.53), and 8.6\% ((0.58-0.53)/0.58), respectively, by applying effort-aware adjustment, \toolname{} outperforms VulFixMiner on every metric (except $P_{opt}@5\%$).
 
From all the aforementioned, we empirically demonstrated that the effort-aware adjustment increases the number of identified vulnerability-fixing commits under specific costs of LOC.

\vspace{2mm}

\begin{table*}[htbp]
\caption{Performance of \toolname{} and VulFixMiner on tangled Java and Python commits}
\label{tab:rq4_all}
\begin{center}
\begin{tabular}{l|l|c|cccc|cccc}
\hline
\textbf{Lang} & \textbf{Model} & \textbf{AUC} & \multicolumn{4}{c|}{\textbf{CostEffort}} & \multicolumn{4}{c}{\textbf{$P_{opt}$}} \\
&  &  & 5\% & 10\% & 15\% & 20\% & 5\% & 10\% & 15\% & 20\% \\
 \hline
\multirow{2}{*}{\textbf{Java}} & \textbf{VulFixMiner} & 0.83 & 0.64 & 0.70 & 0.72 & 0.74 & 0.53 & 0.60 & 0.64 & 0.66 \\
& \textbf{\toolname{}} & \textbf{0.89} & \textbf{0.68} & \textbf{0.80} & \textbf{0.87} & \textbf{0.90} & \textbf{0.56} & \textbf{0.65} & \textbf{0.71} & \textbf{0.76}\\
\hline
\multirow{2}{*}{\textbf{Python}} & \textbf{VulFixMiner} & 0.81 & 0.44 & 0.46 & 0.56 & 0.62 & 0.26 & 0.36 & 0.41 & 0.46 \\
& \textbf{\toolname{}} & \textbf{0.89} & \textbf{0.46} & \textbf{0.62} & \textbf{0.74} & \textbf{0.90} & \textbf{0.28} & \textbf{0.42} & \textbf{0.51} & \textbf{0.58}\\
\hline
\end{tabular}

\end{center}
\end{table*}

\begin{table*}[htbp]
\caption{Regular expression used to filter security-related commits provided by Zhou et al.\cite{zhou2017automated}}
\label{tab:security_regex}
\begin{center}
\resizebox{\textwidth}{!}{
\begin{tabular}{l|c}
\hline
Rule name & Regular Expression \\
\hline
 & \texttt{|(?i)(denial.of.service|\textbackslash bXXE\textbackslash b|remote.code.execution|\textbackslash bopen.redirect|OSVDB|\textbackslash bXSS\textbackslash b|} \\
 
strong\_vuln & \texttt{|\textbackslash bReDoS\textbackslash|\textbackslash bCVE\textbackslash b|\textbackslash bvuln\textbackslash b|\textbackslash bNVD\textbackslash b|malicious|x-frame-−options|attack|cross.site|exploit|} \\

\_patterns & \texttt{|directory.traversal|\textbackslash bRCE\textbackslash b|\textbackslash bdos\textbackslash b|\textbackslash bXSRF\textbackslash b|clickjack|session.fixation|hijack|}\\

 & \texttt{|advisory|insecure|security|\textbackslash bcross-−origin\textbackslash b|unauthori[z|s]ed|infinite.loop)} \\
 
\hline

 & \texttt{|(?i)(authenticat(e|ion)|bruteforce|bypass|constant.time|crack|credential|} \\
 
medium\_vuln & \texttt{|\textbackslash bDoS\textbackslash b|expos(e|ing)|hack|harden|injection|lockout|overflow|password|\textbackslash bPoC\textbackslash b|proof.of.concept|} \\

\_patterns & \texttt{|poison|privelage|\textbackslash b(in)?secur(e|ity)|(de)?serializ|spoof|timing|traversal)} \\
\hline
\end{tabular}
}
\end{center}
\end{table*}

\noindent\textbf{{RQ3. How do different levels of granularity affect the performance of \toolname{}?}}

To answer this RQ, we compare the performance of multiple versions of \toolname{}. First, we have four versions of \toolname{} where in each version, \toolname{} contains only one granularity. Then, starting from one version,  line level as an instance, we continuously integrate more levels of granularity, i.e., hunk-level, file-level, and commit-level until the complete version of \toolname{} is constructed. Note that effort-aware adjustment is applied for every version of \toolname{}.

The performance on the Java and Python dataset is shown in Table \ref{tab:rq3_all}. 
Comparing four versions of \toolname{} that contain single granularity, we can observe that no version clearly outperforms the others across all metrics. However, the complete version of \toolname{} demonstrates the best overall performance. This performance improvement can be attributed to the advantages obtained from combining the different granularities. To further clarify this, we inspect the performance of \toolname{} while incorporating additional granularities on top of the line level. 
For Java, the experimental results in terms of AUC, CostEffort, $P_{opt}$ keep increasing when a new level granularity is added continuously. 
It indicates that all levels of granularity contribute to the performance of \toolname{}.
Specifically, in terms of AUC, \toolname{} improves 4.9\% ((0.85-0.81)/0.81) compared to single level of granularity, i.e., line-level granularity. In terms of CostEffort and $P_{opt}$, the maximum improvements are 8.5\% at CostEffort@10\% and 8\% at $P_{opt}$@5\% respectively. 
For Python, although the experimental results are not linearly increased when each of the levels of granularity is added, the performance of \toolname{} still increased AUC by 2.5\% ((0.83-0.81)/0.81). In terms of effort-aware metrics, the maximum improvements are 6.8\% at CostEffort@5\% and 17.9\% at $P_{opt}$@5\%. Compared to the single level of granularity, i.e., line-level, \toolname{} either outperforms or tie on the remaining thresholds of the effort-related metrics.

Interestingly, \toolname{} utilizing only line-level information even can outperform VulFixMiner on the Python dataset, and demonstrates comparable performance on the Java dataset. This is due to two main reasons. Firstly, breaking down code changes into smaller, more detailed parts allows the deep learning model to consider more meaningful representations by capturing the inter-dependencies between these components, which has been shown to be effective in prior research \cite{hoang2020cc2vec}. Secondly, the gating mechanism of the LSTM enables it to selectively update and retain pertinent information, while disregarding irrelevant information. However, it is important to acknowledge that noise can arise at multiple levels, not solely at the line level as illustrated in our motivating example. Therefore, integrating information from all granularities helps \toolname{} to achieve best performance.

\vspace{2mm}

\noindent\textbf{{RQ4. Can \toolname{} detect vulnerability-fixing commits that involve different types of changes?}}

From the test dataset of Java and Python projects, we extract commits with a large number of disjoint code changes. Specifically, we select commits with five or more hunks. Next, we use the same evaluation metrics in the paper to compare \toolname{} and VulFixMiner in the subsets of Java and Python data. Table \ref{tab:rq4_all} present the experimental results.

Overall, \toolname{} outperforms the state-of-the-art baseline on all the evaluation metrics. On Java, \toolname{} outperforms VulFixMiner by 7.2\% ((0.89-0.83)/0.83) in terms of AUC. Similarly, for Python, \toolname{} outperforms VulFixMiner by 9.9\% ((0.89-0.81)/8.01). In terms of CostEffort@L\%, \toolname{} improved over VulFixMiner by up to 21.6\% ((0.90-0.74)/0.74) and 45.1\% ((0.90-0.62)/0.62). Similarly, $P_{opt}$ reaches the highest improvements at 20\% total LOC, with 15.2\% ((0.76-0.66)/0.66) and 26.1\% ((0.58-0.46)/0.46) on Java and Python respectively.

Combined with the results in Table \ref{tab:rq1_all}, our experiments indicate that \toolname{} has higher discriminative power on commits that have a greater number of hunks. \toolname{} achieves higher AUC on both Java (0.90 versus 0.85) and Python (0.89 versus 0.83). In terms of CostEffort and $P_{opt}$, \toolname{} similarly outperforms VulFixMiner at all thresholds.

%%%%%%%%%%%%%%%%%%%%%%

\begin{table*}[htbp]
\caption{Performance of \toolname{} and VulFixMiner on dataset of security-related commits on Java and Python}
\label{tab:rq6_all}
\begin{center}
\begin{tabular}{l|l|c|cccc|cccc}
\hline
\textbf{Lang} & \textbf{Model} & \textbf{AUC} & \multicolumn{4}{c|}{\textbf{CostEffort}} & \multicolumn{4}{c}{\textbf{$P_{opt}$}} \\
&  &  & 5\% & 10\% & 15\% & 20\% & 5\% & 10\% & 15\% & 20\% \\
 \hline
\multirow{2}{*}{\textbf{Java}} & \textbf{VulFixMiner} & \textbf{0.79}	& 0.27 & 0.34 & 0.47 &	0.54 & 0.17	 & 0.24 & 0.30 & 0.35 \\
& \textbf{\toolname{}} & \textbf{0.79} & \textbf{0.44} & \textbf{0.57} & \textbf{0.66} & \textbf{0.74} & \textbf{0.37} &	\textbf{0.45} &	\textbf{0.51} & \textbf{0.56}\\
\hline
\multirow{2}{*}{\textbf{Python}} & \textbf{VulFixMiner} & 0.67 & 0.25 & 0.27 & 0.36 &	0.40 & 0.16 & 0.22 & 0.27 & 0.29 \\
& \textbf{\toolname{}} & \textbf{0.73} & \textbf{0.50} & \textbf{0.60} & \textbf{0.67} &	\textbf{0.67} & \textbf{0.46} & \textbf{0.51} &	\textbf{0.55} &	\textbf{0.58}\\
\hline
\end{tabular}
\end{center}
\end{table*}

\begin{table*}[htbp]
\caption{Performance of \toolname{} using different PCA settings on Java and Python projects}
\label{tab:rq7_pca_all}
\begin{center}
\begin{tabular}{l|l|c|cccc|cccc}
\hline
\textbf{Lang} & \textbf{Model} & \textbf{AUC} & \multicolumn{4}{c|}{\textbf{CostEffort}} & \multicolumn{4}{c}{\textbf{$P_{opt}$}} \\
&  &  & 5\% & 10\% & 15\% & 20\% & 5\% & 10\% & 15\% & 20\% \\
 \hline
\multirow{6}{*}{\textbf{Java}} & \toolname{}$_{PCA\_{80\%}}$ & 0.25	& 0.11 &	0.15 &	0.22 &	0.25 &	0.06 &	 0.1 &	0.13 &	0.15 \\
& \toolname{}$_{PCA\_{85\%}}$ & 0.63 & 0.50 &	0.62 & 0.70 & 0.75 & 0.36 & 0.46 & 0.53 & 0.58 \\
& \toolname{}$_{PCA\_{90\%}}$ & 0.48 &	0.28 &	0.42 &	0.55 &	0.66 &	0.14 &	0.25 &	0.32 &	0.40 \\
& \toolname{}$_{PCA\_{95\%}}$ & 0.76 & 0.56 &	0.70 &	0.81 & 0.87 & 0.4 & 0.52 & 0.60 & 0.66 \\
& \toolname{}$_{PCA\_{99\%}}$ & 0.83 &	0.62 &	0.76 &	0.83 &	0.87 &	\textbf{0.51} &	\textbf{0.60} & 	\textbf{0.67} &	0.72 \\
& \textbf{\toolname{}} & \textbf{0.85} & \textbf{0.64} & \textbf{0.77} & \textbf{0.87} & \textbf{0.91} & 0.50 & \textbf{0.60} & \textbf{0.67} & \textbf{0.73} \\
\hline
\multirow{6}{*}{\textbf{Python}} & \toolname{}$_{PCA\_{80\%}}$ & 0.27 &	0.06 &	0.13 &	0.17 &	0.2 &	0.03 &	0.06 &	0.09 &	0.12 \\
& \toolname{}$_{PCA\_{85\%}}$ & 0.70	& 0.48 &	0.63 & 0.70 & 0.75 & 0.29 & 0.44 & 0.51 & 0.57 \\
& \toolname{}$_{PCA\_{90\%}}$ & 0.48 & 0.09 &	0.24 &	0.33 &	0.41 &	0.05 &	0.16 &	0.17 &	0.22 \\
& \toolname{}$_{PCA\_{95\%}}$ & 0.75 & 0.36 &	0.48 & 0.61 & 0.72 & 0.25 & 0.33 & 0.41 & 0.47 \\
& \toolname{}$_{PCA\_{99\%}}$ & 0.80 &	0.44 &	0.58 &	0.72 &	0.75 &	0.31 &	0.41 &	0.49 &	0.55 \\
& \textbf{\toolname{}} & \textbf{0.83} & \textbf{0.47} &\textbf{0.64} & \textbf{0.74} & \textbf{0.81} & \textbf{0.33} & \textbf{0.45} & \textbf{0.53} & \textbf{0.59} \\
\hline
\end{tabular}
\end{center}
\end{table*}

%%%%%%%%%%%%%%%%%%%%%%

\begin{table*}[htbp]
\caption{Performance of \toolname{} with and without feature selection (FCBF) on Java and Python projects}
\label{tab:rq7_fcbf_all}
\begin{center}
\begin{tabular}{l|l|c|cccc|cccc}
\hline
\textbf{Lang} & \textbf{Model} & \textbf{AUC} & \multicolumn{4}{c|}{\textbf{CostEffort}} & \multicolumn{4}{c}{\textbf{$P_{opt}$}} \\
&  &  & 5\% & 10\% & 15\% & 20\% & 5\% & 10\% & 15\% & 20\% \\
 \hline
\multirow{2}{*}{\textbf{Java}} & \textbf{\toolname{}$_{FCBF}$} & 0.47 & 0.33 &	0.51 &	0.59 &	0.68 &	0.20 &	0.31 &	0.39 &	0.45 \\
& \textbf{\toolname{}} & \textbf{0.85} & \textbf{0.64} & \textbf{0.77} & \textbf{0.87} & \textbf{0.91} & \textbf{0.50} & \textbf{0.60} & \textbf{0.67} & \textbf{0.73}\\
\hline
\multirow{2}{*}{\textbf{Python}} & \textbf{\toolname{}$_{FCBF}$} & 0.52 & 0.3 & 0.44 &	0.54 &	0.62 &	0.18 &	0.27 &	0.35 &	0.41 \\
& \textbf{\toolname{}} & \textbf{0.83} & \textbf{0.47} & \textbf{0.64} & \textbf{0.74} & \textbf{0.81} & \textbf{0.33} & \textbf{0.45} & \textbf{0.53} & \textbf{0.59}\\
\hline
\end{tabular}

\end{center}
\end{table*}

\section{Discussion}

\subsection{Can \toolname{} distinguish between vulnerability-fixing commits and other type of security-related commits?}
As developers may make changes to secure software, not all security-related commits are vulnerability-fixing. To assess if \toolname{} can distinguish between vulnerability-fixing and other security-related changes, we extract a subset of the data that includes only vulnerability-fixing commits and other types of security-related commits. From both Java and Python test datasets, we extract security-related commits by using the regular expressions (see Table \ref{tab:security_regex}) provided by Zhou et al. \cite{zhou2017automated}. We extract security-related commits from the non-vulnerability-fixing commits by matching them against the regular expressions. A total of 4,023 commits from the Java dataset and 1,455 commits from the Python dataset are extracted. Then, these commits are combined with the 300 vulnerability-fixing commits from the Java dataset, and 195 vulnerability-fixing commits from the  Python dataset.

Table \ref{tab:rq6_all} shows the performance of \toolname{} and VulFixMiner when using only security-related commits for Java and Python, respectively. On both the Java and Python datasets, \toolname{} achieves AUC scores of 0.79 and 0.73. Following Romano et al. \cite{romano2011using}, a classifier with an AUC $\geq$ 0.7 is considered to have achieved acceptable performance. Compared to VulFixMiner, \toolname{} performs equally on the Java dataset with a 0.79 AUC score. On the Python dataset, \toolname{} improves VulFixMiner by 9\% on AUC ((0.73-0.67)/0.67). Regarding CostEffort and $P_{opt}$, \toolname{} outperforms VulFixMiner on every threshold by significant margins. For Java, the improvement varies from 37\% to 63\% and 60\% to 117.6\% on CostEffort and $P_{opt}$, respectively. For Python, the improvement ranges from 67.5\% to 122.2\% and from 100\% to 187.5\% on CostEffort and $P_{opt}$, respectively. Overall, the experimental results indicate that both \toolname{} and VulFixMiner can distinguish vulnerability fixes from other changes to security components.

\subsection{Is there redundancy among the features extracted by \toolname{}?}

\begin{table*}[htbp]
\caption{Performance of \toolname{} when using a different model to extract features for code at one level of granularity for Java and Python projects}
\label{tab:rq8_all}
\begin{center}
\begin{tabular}{l|l|c|cccc|cccc}
\hline
\textbf{Lang} & \textbf{Model} & \textbf{AUC} & \multicolumn{4}{c|}{\textbf{CostEffort}} & \multicolumn{4}{c}{\textbf{$P_{opt}$}} \\
&  &  & 5\% & 10\% & 15\% & 20\% & 5\% & 10\% & 15\% & 20\% \\
 \hline
\multirow{5}{*}{\textbf{Java}} & \toolname{}$_{Line\_{LSTM}}$ & 0.84 &	0.62 &	0.74 &	0.84 &	0.88 &	0.49 &	0.59 & 	0.66 &	0.71 \\
& \toolname{}$_{Line\_{GRU}}$ & \textbf{0.85} & 0.62 &	0.75 &	0.84 &	0.89 &	0.49 &	0.59 &	0.66 &	0.71 \\
& \toolname{}$_{Hunk\_{FCN}}$ & 0.83 &	0.62 &	0.74 &	0.83 &	0.88 &	0.49 &	0.59 &	0.66 &	0.71 \\
& \toolname{}$_{File\_{CNN}}$ & 0.84 &	0.62 &	\textbf{0.77} &	0.84 &	0.90 & 	\textbf{0.50}	& 0.60 & 	0.66 & 	0.72 \\
& \textbf{\toolname{}} & \textbf{0.85} & \textbf{0.64} & 0.77 & \textbf{0.87} & \textbf{0.91} & \textbf{0.50} & \textbf{0.60} & \textbf{0.67} & \textbf{0.73} \\
\hline
\multirow{5}{*}{\textbf{Python}} & \toolname{}$_{Line\_{LSTM}}$ & 0.81 &	0.42 &	0.64 &	0.7 &	0.78 &	0.28 &	0.41 &	0.5 & 	0.56 \\
& \toolname{}$_{Line\_{GRU}}$ & 0.81 &	0.43 &	0.63 &	0.69 &	0.79 &	0.28 &	0.41 &	0.49 & 	0.56 \\
& \toolname{}$_{Hunk\_{FCN}}$ & 0.81 &	0.46 &	0.61 &	0.71 &	0.78 &	0.29 &	0.42 &	0.50 &	0.56 \\
& \toolname{}$_{File\_{CNN}}$ & 0.82 &	\textbf{0.51} &	\textbf{0.66} &	0.70 &	0.79 &	0.32 &	0.45 &	\textbf{0.53} &	0.58 \\
& \textbf{\toolname{}} & \textbf{0.83}	& 0.47 &	0.64 &	\textbf{0.74} &	\textbf{0.81} &	\textbf{0.33} &	\textbf{0.45} &	\textbf{0.53} &	\textbf{0.59} \\
\hline
\end{tabular}

\end{center}
\end{table*}

To understand the importance of the extracted features, we compare the performance of \toolname{} with and without applying feature reduction, Principal Composition Analysis - PCA\cite{wold1987principal}, or feature selection technique (Fast Correlation-based Feature Selection - FCBF\cite{yu2003feature})

\textbf{\textit{\toolname{} with PCA}}. To study the redundancy of features, PCA has been applied in different studies, including software engineering\cite{massoudi2021software,pandey2020software,goel2020defect}. Similarly, in our case, it can be used to reduce the feature space of the input vector to the neural classifier. Specifically, after obtaining the features from different granularities, we use Principal Component Analysis (PCA) to obtain the principal components and use them as inputs for the neural classifier. If the principal components obtain the same performance as the original feature vectors, it implies that some original features were redundant.  

PCA computes new features called principal components, obtained from linear combinations of the original features\cite{abdi2010principal}. PCA obtains these features by projecting the original features onto a lower dimensional space such that the variance of the projected data is maximized. The principal components are computed such that the first principal component will explain the most variance in the dataset, followed by the second component, and so on \cite{abdi2010principal}. 
Hence, to assess if feature vectors extracted by different granularities are important for \toolname{}, we concatenate them into a vector, and we perform PCA on the combined vector before passing the principal components to the neural classifier. Following Kondo et al.\cite{kondo2019impact}, PCA is configured so that it explains a specific proportion of variance in the data. In our experiments, we opt to retain 80\%, 85\%, 90\%, 95\%, and 99\% of the variance in the data, respectively.

Table \ref{tab:rq7_pca_all} illustrates the performance of \toolname{} in these cases. 
As we can see, \toolname{} performs worse using the principal components. Without PCA, \toolname{} achieves the highest scores in every evaluation metric on both Java and Python (except $P_{opt}$@5\% on Java, with a marginal 0.01 decrease). Thus, feature selection does not help increase the performance of \toolname{}.

\textbf{\toolname{} with FCBF}. Fast Correlation-based Feature Selection (FCBF) \cite{yu2003feature} is a feature selection technique, which has been shown to be effective in removing  redundant features  in different tasks\cite{auld2007bayesian, nguyen2012heterogeneous, li2009efficient, kannan2010novel}. Unlike feature reduction techniques, which compute a new set of features, feature selection techniques such as FCBF  selects the most important features to be retained, removing other features. Similar to PCA, we apply FCBF after concatenating all feature vectors from different granularities. Then the output of the FCBF is passed as the input to the neural classifier.

Table \ref{tab:rq7_fcbf_all} illustrates the performance of \toolname{} with and without using FCBF on Java and Python. After applying FCBF, the performance of \toolname{} is reduced on every evaluation metric. For example, by applying FCBF, the AUC scores drop by 45\% ((0.85-0.47)/0.85) and 37\% ((0.83-0.52)/0.83 on Java and Python, respectively. The results suggest that \toolname{} does not benefit from feature selection. As neither feature reduction nor feature selection improves the performance of \toolname{}, we conclude that there is a low level of redundancy among the features extracted by \toolname{}. 

\subsection{Does the choice of neural network for feature extractor affect the performance of \toolname{}?}

As described in Section \ref{sec:feature_extractor}, \toolname{} uses different deep learning models to extract code features at different granularity levels. Therefore, we perform a set of experiments to observe the performance of \toolname{} when using different deep learning models for extracting features. In each experiment, we replace the current feature extractor model at one granularity with another design. Specifically, for line-level granularity, we replace our design BiLSTM with either LSTM or GRU. We denote the two corresponding versions of \toolname{} when using these two models at line level granularity as \toolname{}$_{Line\_{LSTM}}$ and \toolname{}$_{Line\_{GRU}}$, respectively. Similarly, for hunk-level granularity, we replace CNN with FCN, and for file-level granularity, we replace FCN with CNN. The replacements yield two other versions of \toolname{}, namely \toolname{}$_{Hunk\_{FCN}}$ and \toolname{}$_{File\_{CNN}}$. Table \ref{tab:rq8_all} shows the performance of \toolname{} for Java and Python when using different neural network models for a level of granularity. 

Compared to the variants of \toolname{} where the feature extractor for one level of granularity uses a different model, \toolname{} achieves the highest AUC on both Java and Python, with higher scores of either 0.01 or 0.02. However, on the effort-aware metrics, \toolname{}$_{File\_{CNN}}$ outperforms \toolname{} on CostEffort@5\% and CostEffort@10\% on the Python dataset by 8.5\% and 3.1\%, respectively. Overall, we see that different model designs slightly affect the performance of \toolname{}. Nevertheless, when using the proposed design in Section 4.5.1, \toolname{} achieves the highest results on most evaluation metrics. It confirms our intuition in designing the feature extractors.
\subsection{How does \toolname{} perform in different contexts of inspection cost?}

\begin{table}[th]
\caption{Performance of \toolname{} and VulFixMiner for the Java and Python projects on CostEffort@L\% Hunk level}
\label{tab:dis1_all_hunk}
\begin{center}
\begin{tabular}{l|l|cccc}
\hline
\textbf{Lang} & \textbf{Model} & \multicolumn{4}{|c}{\textbf{CostEffort\_Hunk}} \\
&  & 5\% & 10\% & 15\% & 20\% \\
 \hline
\multirow{2}{*}{\textbf{Java}} & \textbf{VulFixMiner} & \textbf{0.54} & \textbf{0.63} & 0.66 & 0.69 \\
& \textbf{\toolname{}} & \textbf{0.54} & \textbf{0.63} & \textbf{0.67} & \textbf{0.72} \\
\hline
\multirow{2}{*}{\textbf{Python}} & \textbf{VulFixMiner} & 0.31 & 0.39 & 0.46 & 0.53 \\
& \textbf{\toolname{}} & \textbf{0.46} & \textbf{0.63} & \textbf{0.72} & \textbf{0.79} \\
\hline
\end{tabular}

\end{center}
\end{table}

\begin{table}[th]
\caption{Performance of \toolname{} and VulFixMiner for the Java and Python projects on CostEffort@L\% File level}
\label{tab:dis1_all_file}
\begin{center}
\begin{tabular}{l|l|cccc}
\hline
\textbf{Lang} & \textbf{Model} & \multicolumn{4}{|c}{\textbf{CostEffort\_File}} \\
&  & 5\% & 10\% & 15\% & 20\% \\
 \hline
\multirow{2}{*}{\textbf{Java}} &  \textbf{VulFixMiner} & \textbf{0.58} & 0.64 & 0.67 & 0.70 \\
& \textbf{\toolname{}} & 0.57 & \textbf{0.67} & \textbf{0.75} & \textbf{0.81} \\
\hline
\multirow{2}{*}{\textbf{Python}} & \textbf{VulFixMiner} & 0.30 & 0.37 & 0.45 & 0.50 \\
& \textbf{\toolname{}} & \textbf{0.43} & \textbf{0.58} & \textbf{0.65} & \textbf{0.72} \\
\hline
\end{tabular}

\end{center}
\end{table}

\begin{table}[h!]
\caption{Performance of \toolname{} and VulFixMiner for the Java and Python projects on CostEffort@L\% Commit level}
\label{tab:dis1_all_commit}
\begin{center}
\begin{tabular}{l|l|cccc}
\hline
\textbf{Lang} & \textbf{Model} & \multicolumn{4}{|c}{\textbf{CostEffort\_Commit}} \\
 & & 5\% & 10\% & 15\% & 20\% \\
 \hline
\multirow{2}{*}{\textbf{Java}} & \textbf{VulFixMiner} & \textbf{0.54} & \textbf{0.63} & 0.66 & 0.69 \\
& \textbf{\toolname{}} & \textbf{0.54} & \textbf{0.63} & \textbf{0.67} & \textbf{0.72} \\
\hline
\multirow{2}{*}{\textbf{Python}} & \textbf{VulFixMiner} & 0.28 & 0.37 & 0.44 & 0.50 \\
& \textbf{\toolname{}} & \textbf{0.41} & \textbf{0.57} & \textbf{0.63} & \textbf{0.68} \\
\hline
\end{tabular}

\end{center}
\end{table}

As the current effort-aware metrics uses LOC as a measure of the inspection effort, we are curious about the performance of \toolname{} and the state-of-the-art baseline, VulFixMiner, when using other measures of effort, e.g., the number of hunks, files, commits inspected.
Specifically, similar to CostEffort@L\% described in Section \ref{sec:metrics}, we defined CostEffort\_Hunk@L\%, CostEffort\_File@L\%, CostEffort\_Commit@L\% which are the CostEffort calculated using the number of inspected hunks, files, commits respectively. The results are illustrated in Tables \ref{tab:dis1_all_hunk}  \ref{tab:dis1_all_file},
\ref{tab:dis1_all_commit}. Combined with the results in Tables \ref{tab:rq1_all}, our experimental results show that on four measures, LOC, hunk, file, and commit, \toolname{} either outperforms VulFixMiner or performs similarly. This validates our findings from before that \toolname{} leads to a reduction in effort compared to VulFixMiner.

\begin{table}[]
\caption{Number of detected large vulnerability-fixing commits of \toolname{} and VulFixMiner for Java and Python projects}
\label{tab:dis2_all_large}
\begin{center}
\begin{tabular}{l|l|cccc}
\hline
\textbf{Lang} & \textbf{Model} & \multicolumn{4}{|c}{\textbf{Inspection Cost}} \\
 & & 5\% & 10\% & 15\% & 20\% \\
 \hline
\multirow{3}{*}{\textbf{Java}} & \textbf{VulFixMiner} & 85 & 96 & 99 & 101 \\
& \textbf{\toolname{}} & \textbf{89} & \textbf{100} & \textbf{109} & \textbf{114} \\
\hline
& \textbf{Total No. VF commits} & \multicolumn{4}{|c}{131} \\
\hline
\multirow{3}{*}{\textbf{Python}} & \textbf{VulFixMiner} & 8 & 10 & 12 & 14 \\
& \textbf{\toolname{}} & \textbf{11} & \textbf{14} & \textbf{15} & \textbf{15} \\
\hline
& \textbf{Total No. VF commits} & \multicolumn{4}{|c}{26} \\
\hline
\end{tabular}

\end{center}
\end{table}

\begin{table}[h]
\caption{Number of detected small vulnerability-fixing commits of \toolname{} and VulFixMiner for Java and Python projects}
\label{tab:dis2_all_small}
\begin{center}
\begin{tabular}{l|l|cccc}
\hline
\textbf{Lang} & \textbf{Model} & \multicolumn{4}{|c}{\textbf{Inspection Cost}} \\
 & & 5\% & 10\% & 15\% & 20\% \\
 \hline
\multirow{3}{*}{\textbf{Java}} & \textbf{VulFixMiner} & \textbf{11} & 11 & 11 & 11 \\
& \textbf{\toolname{}} & 10 & \textbf{14} & \textbf{14} & \textbf{15} \\
\hline
& \textbf{Total No. VF commits} & \multicolumn{4}{|c}{15} \\
\hline
\multirow{3}{*}{\textbf{Python}} & \textbf{VulFixMiner} & 9 & 13 & 15 & 15 \\
& \textbf{\toolname{}} & \textbf{12} & \textbf{17} & \textbf{20} & \textbf{26} \\
\hline
& \textbf{Total No. VF commits} & \multicolumn{4}{|c}{30} \\
\hline
\end{tabular}

\end{center}
\end{table}

\subsection{How does \toolname{} perform on large/small datapoints?}

%%%%%%%%%%

We study the effect of commit size on the performance of \toolname{}. In particular, we investigate the performance of \toolname{} on large and small commits. We consider commits that exceed the limit of CodeBERT, i.e., 512 tokens,  as large commits. For small commits, we selected code changes with less than 50 tokens. Our experimental results are illustrated in Tables \ref{tab:dis2_all_large}, \ref{tab:dis2_all_small}.

From the tables, across the programming languages, size, and inspection cost settings, \toolname{} outperforms VulFixMiner on 15 out of the 16 settings. The result shows that \toolname{} can detect vulnerability-fixing commits even when the commits are large or small, and does so better than VulFixMiner.

\subsection{In terms of effort-aware metrics, why does \toolname{} perform better on Java compared to Python despite the same AUC?}

As shown in Table \ref{tab:rq1_all}, while \toolname{} achieves similar AUC scores on the Java and Python datasets (0.85 versus 0.83), there is a considerable difference between the performance of \toolname{} in effort-aware metrics (i.e., CostEffort, $P_{opt}$). This shows that \toolname{} can be considered to be more effective on the Java dataset despite having the same predictive power on both datasets. We investigate further to shed more light on this result by analyzing the number of commits inspected at each effort threshold, which influences the computation of CostEffort and $P_{opt}$.  

Table \ref{tab:dis3} provides the percentage of commits that are inspected from the Java and Python datasets for each effort threshold. At every considered threshold, the proportion of inspected commits from the Java dataset is higher than in Python. This implies that while the predictive power of \toolname{} is similar on both datasets, the number of commits inspected under the same effort thresholds is different. As more commits are inspected on the Java dataset, a greater proportion of vulnerability-fixing commits would be detected using the same amount of effort. Hence, \toolname{} achieves a higher CostEffort and $P_{opt}$ on the Java dataset.

\begin{table}[t]
\caption{Percentage of inspected commits by \toolname{} based on \%LOC}
\label{tab:dis3}
\begin{center}
\begin{tabular}{l|c|c|c|c}
\hline
\textbf{Dataset} & \textbf{5\%LOC} & \textbf{10\%LOC} & \textbf{15\%LOC} & \textbf{20\%LOC} \\
 \hline
\textbf{Java} & 10.9 & 25.5 & 38.4 & 48.8 \\
\textbf{Python} & 7.2 & 15.7 & 25.2 & 34 \\
\hline
\end{tabular}

\end{center}
\end{table}

\begin{table}[t]
\caption{Statistics of Zhou et al. \cite{zhoufinding} dataset following time-aware setting. We refer to vulnerability-fixing commits and non-vulnerability-fixing commits as V.F. and N.V.F, respectively.}
\label{tab:dis_4_stats}
\begin{center}
\begin{tabular}{c|c|c|c|c|c|c}
\hline
% \cline{2-4} 
& \multicolumn{2}{c|}{\textbf{Training Set}} & \multicolumn{2}{c|}{\textbf{Validation Set}} & \multicolumn{2}{c}{\textbf{Testing Set}} \\

& \#V.F. & \#N.V.F. & \#V.F. & \#N.V.F  & \#V.F. & \#N.V.F.  \\
\hline

\textbf{Java} & 979 & 105,158 & 110 & 3,364 & 270 & 13,150 \\

\textbf{Python} & 548 & 69,480 & 61 & 3,154 & 152 & 5,375 \\
\hline
\end{tabular}\\ \vspace{1mm}
V.F.: Vulnerability-fixing Commits, N.V.F.: Non-vulnerability-fixing Commits.

\end{center}
\end{table}

\subsection{\toolname{} under time-aware constraint}
In the original setting (Section \ref{sec:evaluation}), our dataset is split in project-wise manner. In this case, the commits in test data are from different projects which are never seen in the training and validation sets. Nonetheless, the problem of identifying vulnerability-fixing commits can be viewed from a different perspective, i.e., how vulnerability-fixing commits evolve with time. Specifically, can a trained model of \toolname{}, which is based on historical vulnerability-fixing commits, correctly identify future ones? To answer it, we first sorted all the commits in the dataset in chronological order, following a study by Feargus et al. \cite{pendlebury2019tesseract}. Additionally, to ensure that the training and validation data contained an adequate number of vulnerability-fixing commits, we incrementally collected data until we reached the point where 80\% of the collected commits were vulnerability-fixing commits. As a result, we produced a dataset whose statistics are described in Table \ref{tab:dis_4_stats}.

Subsequent to training \toolname{} and the best baseline, VulFixMiner, on the new dataset, we evaluate their performance in terms of AUC, CostEffort, and $P_{opt}$. The result of the evaluation is shown in Table \ref{tab:dis_4_all}.  The outcome indicates that \toolname{} continues to outperform VulFixMiner on both Java and Python on all evaluation metrics. 

%%%%%%%%%%%%%%%%%%%%%%%%%%%%%%

\begin{table*}[htbp]
\caption{Performance of \toolname{} and VulFixMiner on Zhou et al. \cite{zhoufinding} dataset following time-aware setting.
}
\label{tab:dis_4_all}
\begin{center}
\begin{tabular}{l|l|c|cccc|cccc}
\hline
\textbf{Lang} & \textbf{Model} & \textbf{AUC} & \multicolumn{4}{c|}{\textbf{CostEffort}} & \multicolumn{4}{c}{\textbf{$P_{opt}$}} \\
&  &  & 5\% & 10\% & 15\% & 20\% & 5\% & 10\% & 15\% & 20\% \\
 \hline
\multirow{2}{*}{\textbf{Java}} & \textbf{VulFixMiner} & 0.76 & 0.36 & 0.45 & 0.52 & 0.57 & 0.32 & 0.36 & 0.41 & 0.44 \\
& \textbf{\toolname{}} & 0.77 & 0.50 & 0.63 & 0.74 & 0.78 & 0.40 & 0.49 & 0.56 & 0.61 \\
\hline
\multirow{2}{*}{\textbf{Python}} & \textbf{VulFixMiner} & 0.73 & 0.26 & 0.39 & 0.44 & 0.47 & 0.19 & 0.27 & 0.31 & 0.35 \\
& \textbf{\toolname{}} & 0.77 & 0.36 & 0.53 & 0.63 & 0.72 & 0.29 & 0.38 & 0.45 & 0.51 \\
\hline
\end{tabular}

\end{center}
\end{table*}

\subsection{Applicability of \toolname{} on monitoring real project}
In this discussion, we aim to assess the generalizability and applicability of \toolname{} on monitoring a real project. To do so, we run experiments on a new dataset collected from the TensorFlow\footnote{\url{https://github.com/tensorflow/tensorflow}} framework, which is not part of the Zhou et al. \cite{zhoufinding} dataset. 

We first collected vulnerability-fixing commits associated with vulnerabilities reported in the National Vulnerability Database (NVD) from 18 September 2020 to 8 January 2022, and excluded all commits whose messages contained security-related keywords proposed by Zhou et al. \cite{zhoufinding}. It resulted in a total of 284 vulnerability-fixing commits.

Next, we collected all commits of TensorFlow within the same time frame and considered them as non-vulnerability-fixing commits. Note that we exclusively considered source code files written in C or Python, as they are the primary programming languages for TensorFlow. To minimize the impact of large commits, we applied an approach similar to Zhou et al. \cite{zhoufinding} by establishing two thresholds using the 95th percentile of the total modified lines of code (310) and the number of changed files (7) of vulnerability fixes. As a result, we obtained 284 vulnerability-fixing and 16,083 non-vulnerability-fixing commits. We then chronologically split the dataset into training and testing sets in an 80:20 ratio. The setting is consistent with a real-life scenario in which vulnerability-fixing classification models are trained on historical commits and deployed to predict new ones, and mitigate time and spatial bias \cite{pendlebury2019tesseract, lyu2023chronos}, in our evaluation. Finally, we obtained a total of 200 vulnerability-fixing commits and 13,155 non-vulnerability-fixing commits for the training set. For the testing set, we obtained 84 vulnerability-fixing commits and 2,928 non-vulnerability-fixing commits.

\begin{table}[t]
\caption{Performance of \toolname{} on TensorFlow dataset}
\label{tab:dis_5}
\begin{center}
\begin{tabular}{c|cccc|cccc}
\hline
\textbf{AUC} & \multicolumn{4}{c|}{\textbf{CostEffort}} & \multicolumn{4}{c}{\textbf{$P_{opt}$}} \\
& 5\% & 10\% & 15\% & 20\% & 5\% & 10\% & 15\% & 20\% \\
 \hline
0.88 & 0.81 & 0.92 & 0.93 & 0.94 & 0.58 & 0.74 & 0.80 & 0.84 \\
\hline
\end{tabular}

\end{center}
\end{table}

Table \ref{tab:dis_5} shows the performance of \toolname{} on the TensorFlow dataset. It can be seen that \toolname{} can effectively classify vulnerability-fixing commits in the TensorFlow framework with an AUC of 0.88. The results are comparable to its performance on Java and Python datasets which are 0.85 and 0.83, respectively, as reported in the Table 3. The results indicate that \toolname{} can generalize over different programming languages and projects. Moreover, these experimental results also suggest that \toolname{} can significantly reduce human efforts in identifying vulnerability-fixing commits. For instance, \toolname{} can detect 81\% of vulnerabilities by examining just 5\% of the lines of code, and this figure increases to 94\% when examining 20\% of the code. The results show that \toolname{} is promising on reducing human efforts on monitoring vulnerability-fixing commits from a real project, i.e, TensorFlow.

Despite its effectiveness in reducing the human effort required for monitoring vulnerability-fixing commits, MiDas still necessitates the involvement of security experts to validate the presence of such commits. This process can be prone to errors and requires security experts to have a thorough understanding of the codebase to verify these commits. Consequently, this poses a challenge when using MiDas to monitor vulnerability-fixing commits in real-world scenarios. However, it is worth noting that this challenge is not unique to our tool and the identification of vulnerability-fixing commits. Many other software engineering tasks face similar challenges, such as automated program repair~\cite{le2019reliability}, bug detection~\cite{kharkar2022learning} or vulnerability detection~\cite{le2022finding}. Addressing these challenges falls beyond the scope of our current paper, and we leave them as potential directions for future research and investigation.

\subsection{Is ensemble learning needed?}

\begin{figure}
    \centering
    \includegraphics[width=0.4\textwidth]{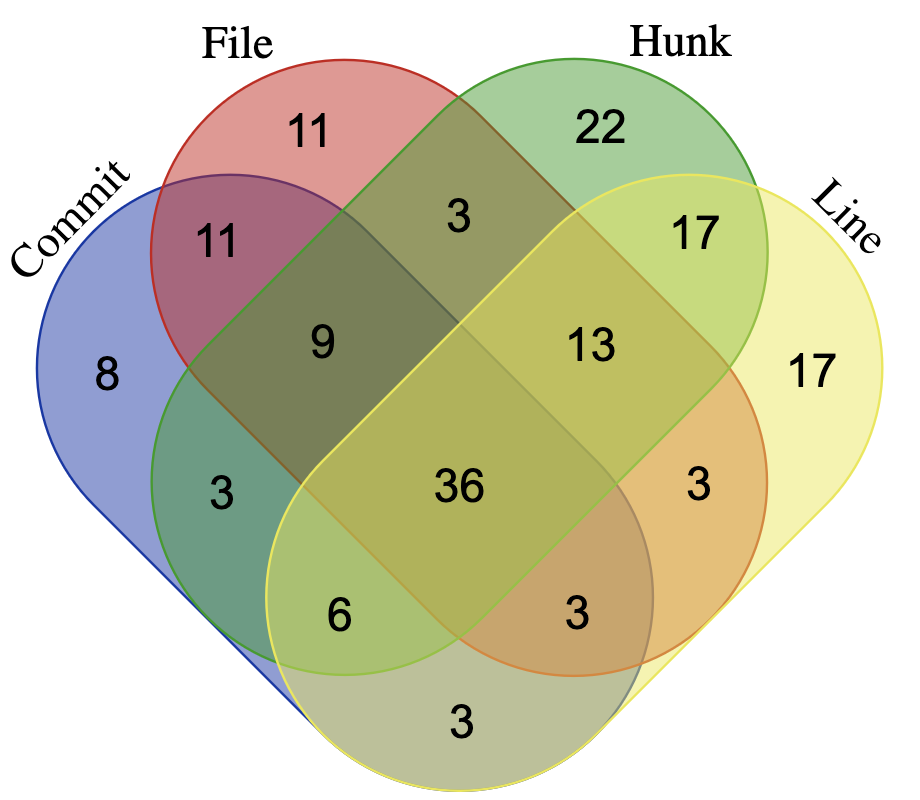}
    \caption{Intersection of correctly detected vulnerability-fixing commits from different granularities in \toolname{}}
    \label{fig:dis_6}
\end{figure}

\begin{table*}[htbp]
\caption{Mean and Standard Deviation (Stdev) in performance of \toolname{} and VulFixMiner for running RQ1 20 times
}
\label{tab:dis_7_stats}
\begin{center}
\begin{tabular}{l|l|c|c|cccc|cccc}
\hline
\textbf{Lang} & \textbf{Model} & \textbf{Stat} &  \textbf{AUC} & \multicolumn{4}{c|}{\textbf{CostEffort}} & \multicolumn{4}{c}{\textbf{$P_{opt}$}} \\
&  &  &  & 5\% & 10\% & 15\% & 20\% & 5\% & 10\% & 15\% & 20\% \\
 \hline
\multirow{4}{*}{\textbf{Java}} & \multirow{2}{*}{\textbf{VulFixMiner}} & Mean & 0.79 & 0.57 & 0.62 & 0.66 & 0.69 & 0.50 & 0.55 & 0.58 & 0.60 \\
 & & Stdev &  0.018 & 0.020 & 0.023 & 0.021 & 0.022 & 0.014 & 0.014 & 0.017 & 0.018 \\

& \multirow{2}{*}{\textbf{\toolname{}}} & Mean & 0.84 & 0.63 & 0.76 & 0.84 & 0.88 & 0.50 & 0.60 & 0.67 & 0.72 \\
 & & Stdev & 0.003 & 0.007 & 0.010 & 0.009 & 0.007 & 0.004 & 0.004 & 0.006 & 0.005 \\
\hline
\multirow{4}{*}{\textbf{Python}} & \multirow{2}{*}{\textbf{VulFixMiner}} & Mean & 0.71 & 0.31 & 0.41 & 0.50 & 0.56 & 0.22 & 0.30 & 0.35 & 0.40 \\
& & Stdev & 0.028 & 0.033 & 0.039 & 0.040 & 0.051 & 0.021 & 0.026 & 0.029 & 0.032 \\

& \multirow{2}{*}{\textbf{\toolname{}}} & Mean & 0.81 & 0.44 & 0.61 & 0.72 & 0.79 & 0.29 & 0.42 & 0.50 & 0.56 \\
& & Stdev & 0.005 & 0.016 & 0.015 & 0.014 & 0.014 & 0.006 & 0.008 & 0.007 & 0.006 \\
\hline
\end{tabular}

\end{center}
\end{table*}

Our approach \toolname{} employs ensemble learning. To assess the need for ensembling, we investigated whether there are unique vulnerability-fixing commits that only a specific granularity can detect. Using a threshold of 0.5, we counted the number of commits that each granularity exclusively detects, as shown in Figure \ref{fig:dis_6}. The figure depicts the intersection of correctly detected vulnerability-fixing commits from classifiers corresponding to different granularities. 
We can see that these classifiers only agree on 36 out of 165 vulnerability-fixing commits, accounting for less than 22\% of the commits detected by all granularities. Furthermore, Figure \ref{fig:dis_6} shows that 8, 11, 22, and 17 commits can only be detected by the Commit level, File level, Hunk level, and Line level, respectively. These results demonstrate that information from different granularities is useful for detecting different vulnerability-fixing commits. Therefore, combining information from different granularities can improve the performance of vulnerability-fixing commit classification. Indeed, \toolname{} achieves the best performance by using all granularities as we have shown in RQ3.

\subsection{Effects of randomness on the performance of \toolname{}}
To address the potential impact of randomness in our deep learning experiments, we ran \toolname{} and VulFixMiner 20 times using different seeds for each version of both models and calculated the average results and standard deviations, which are reported in Table \ref{tab:dis_7_stats}. The results demonstrate that both \toolname{} and VulFixMiner are stable, with a standard deviation of less than 0.05. 

Furthermore, to check the statistical significance of our findings, we utilized the Wilcoxon Signed-Rank Test\cite{wilcoxon1992individual}, which is a non-parametric hypothesis test commonly used in previous studies \cite{yan2019characterizing, yan2020just, yatish2019mining}, at a 95\% confidence level. For each evaluation metric, we set the null hypothesis that there is no difference between the performance of \toolname{} and VulFixMiner, and calculated the corresponding p-value. If the p-value is less than 0.05, we reject the null hypothesis and conclude that \toolname{} outperforms VulFixMiner. We find that except for $P_{opt}$@5\% on the Java dataset, all other metrics on the Java and Python datasets had p-values of less than 0.05. Therefore, we conclude that \toolname{} statistically significantly outperforms VulFixMiner on all metrics except Popt@5\% for the Java dataset.

\section{Threats to Validity}
\label{sec:threats}

\textbf{Threats to internal validity} relates to the mistakes in the implementation and analysis of \toolname{}. To mitigate the threats, we have double-checked our source code and data. In our experiments, we used the same CodeBERT version \cite{codebert_base} for every base model to ensure there is no difference between the used pre-trained models. Moreover, all code fragments after extracted are represented by CodeBERT in the same way as we proposed in Section \ref{sec:approach}.
Our source code and data are available in our replication package~\cite{replication_package}, which future work can analyze and build on.

To minimize the threats to  \textbf{construct validity}, we used the standard evaluation metrics, which have been used in numerous studies in software engineering.
For a fair comparison, we also used exactly the same dataset with the same separation as the prior study \cite{zhoufinding} for our comparison with the baseline models. 

To reduce the risk from threats to \textbf{external validity}, which is related to the generalizability of our findings, the dataset used in our experiments contains a large number of commits from a wide range of projects. Besides, the datasets cover two popular program languages, Java and Python. The result shows that our representations for code changes can work in both Java and Python. Additionally, we run \toolname{} on the TensorFlow dataset, which contains code changes in C and collected from National Vulnerability Database (NVD). The experiment reduces the risk that \toolname{} can only work on an artificial dataset.
However, the performance of \toolname{} may not be generalized to other programming languages. As we use CodeBERT to represent code fragments, the quality of the representation is affected  by the CodeBERT pretrained model. Specifically, CodeBERT is pre-trained on only six programming languages. It may, therefore, have limitations in representing code from other  programming languages. However, according to a recent study by Chen et al. \cite{chen2022transferability}, language model pre-trained on high-resource programming languages (e.g., Java and Python) can be generalized even to low-resource programming languages. Depending on the downstream tasks, to achieve high performance, selecting a suitable programming language for finetuning is crucial. Therefore, we leave the analysis of \toolname{} on other programming languages for future work.

% \giang{
% However, the performance of \toolname{} may not be generalized to other programming languages. As we use CodeBERT to represent code fragment, the quality of the representation is related to CodeBERT pretrained model. Specifically, CodeBERT, a multilingual pre-trained language model (PLM), is pre-trained on six programming languages only, therefore, does not present all languages. According to Chen et al.\cite{chen2022transferability}, understanding if PLM pre-trained on high-resource programming languages (e.g. Java and Python) can be utilized for other programming languages is important. Moreover, depending on the downstream tasks, to achieve high performance, selecting suitable programming languages for fine-tuning is crucial.
% }
As our last point to reduce the risk, the proposed effort-aware objective function is applicable for different datasets, where the distribution of commits' length may not be the same as the one we used in our experiments.

\section{Related Work}
\label{sec:related}

Many works have been proposed to identify vulnerability-fixing commits based on both commit messages and code changes, e.g., \cite{sabetta2018practical, zhou2021spi}. Sabetta et al.\cite{sabetta2018practical} trained two linear Support Vector Machine models based on Bag-Of-Words representation for classifying commit message and code change, respectively. For each commit, the predictions of the two classifiers are combined using a simple voting mechanism to flag if a commit is for vulnerability-fixing or not. Zhou et al.\cite{zhou2021spi} leverage LSTM and multi-layer CNN to train a commit message classifier and a code change classifier, then the results from the two classifiers are combined by using a stacking ensemble. Nguyen et al.\cite{nguyen2022hermes} further consider the commit issue as an additional source of information for classification. Their model is an extension of the work from Sabetta et al.\cite{sabetta2018practical}, with adding a new component of the commit issue classifier. Different from these studies, following the practice where vulnerability-related information should not be explicitly mentioned, \toolname{}  considers only code changes to identify vulnerability-fixing commits.

There also exist some works focus on other types of commit-related classification problems. For example, VCCFinder\cite{perl2015vccfinder} utilizes an SVM-based model based on hand-craft features to identify vulnerability-introducing commits. These features come from different scopes, i.e., project, author, commit, and file. DeepCVA\cite{le2021deepcva} adapts multi-task learning technique to tackle the problem of characterizing vulnerability-introducing commits to provide timely information about the exploitability, impact and severity of the vulnerabilities. They use attention-based convolutional gated recurrent units to extract code change and its surrounding context within a vulnerability-introducing commit. DEPA~\cite{zhong2020inferring} utilizes the partial-code analysis tool GRAPA~\cite{zhong2017boosting} to analyze previous bug-fixing commits, extracting bug signatures and then employing the signatures to detect new bugs.
Our work is similar in that we analyze commits to understand buggy patterns and security risks, however, our focus is vulnerability-fixing commits. 

Apart from vulnerability-related commits identification, other studies have proposed methods of classifying commits based on different categorizations. DeepJIT\cite{hoang2019deepjit} is built upon CNN to represent commit message and code changes features. Features of the two sources of information are combined by a fully connected network to predict Just-in-time defects. CC2Vec\cite{hoang2020cc2vec} is evaluated on the same task, however, only considers code changes. The core of CC2Vec is the Hierarchical Attention Network used to extract code change's feature and a set of comparison functions for capturing the difference between removed code and added code. 
Subsequently, LApredict\cite{zeng2021deep} empirically demonstrates that a Logistic Regression model with only one feature, i.e., added-line-number, can outperform deep learning in just-in-time defect prediction. 

However, the vulnerability-fixing detection task dataset is more imbalanced compared to the just-in-time defect prediction task dataset, mainly due to the limited number of existing vulnerability-fixing commits. Specifically, only 0.34\% of commits in our dataset are vulnerability-fixing commits, whereas defective commits make up between 8.64\% and 41.20\% of all commits in the LApredict datasets \cite{zeng2021deep}. Moreover, the aforementioned approaches \cite{zeng2021deep, hoang2019deepjit, kochhar2015code} solely rely on code change information at a single granularity. In contrast, our study proposes an ensemble learning framework that utilizes code change information across multiple granularities to mitigate the impact of data imbalance. Ensemble learning has been proven to be effective to alleviate the data imbalance problem \cite{haixiang2017learning, khoshgoftaar2015ensemble, dong2020survey} by combining multiple base models and training them separately.

The most related studies to our work are VulFixMiner~\cite{zhoufinding} and CoLeFunDa~\cite{zhoucolefunda} which are also aiming to classify vulnerability-fixing commits only based on code changes. Similar to our approach, VulFixMiner and CoLeFunDa also use CodeBERT\cite{feng2020codebert} to represent code changes. However, they only consider single-level (i.e., either file-level or function-level) granularity for the representation while \toolname{} considers multiple granularities of a code change to precisely capture fix-related information and untangle it from noise. This enables \toolname{} to outperform VulFixMiner in terms of discriminative ability.
% The most related study to our work is VulFixMiner\cite{zhoufinding} which is also aiming to classify vulnerability-fixing commits only based on code changes. Similar to our approach, VulFixMiner also uses CodeBERT\cite{feng2020codebert} to represent code changes. However, it only considers single-level (i.e., file-level) granularity for the representation while \toolname{} considers multiple granularities of a code change to precisely capture fix-related information and untangle it from noise. This enables \toolname{} to outperform VulFixMiner in terms of discriminative ability.
Moreover, \toolname{} uses an effort-aware adjustment function to further boost the performance of \toolname{} in reducing the amount of effort for inspecting the commits.
Since CoLeFunDa has not released a replication package, we cannot compare CoLeFunDa to our work.

\section{Conclusion and Future Work}
\label{sec:conclusion}
In this paper, we propose \toolname{}, a multi-granularity deep learning model for vulnerability-fixing commit detection. Our findings suggest that representing commit code changes in different levels of granularity could effectively improve the performance compared to the state-of-the-art baseline. Moreover, we take the effort-aware evaluation metrics into consideration to evaluate approaches in the real-world scenario. According to the result, the proposed effort-aware adjustment function has demonstrated its effectiveness of reducing the inspection cost of developers in detecting vulnerability-fixing commits. 

The current version of \toolname{} leverages only source code information in code changes to detect vulnerability-fixing commits. In the future, we will explore other sources of information such as code comments, project-related data, etc.  to further improve our model. 
Besides, we plan to investigate the impact of different kinds of effort-aware adjustment functions on overall performance of \toolname{}.
Moreover, we plan to enhance the representation of code changes by incorporating additional context through dependent code analysis techniques, such as data-flow analysis. 
We also plan to include semantic analysis which allows \toolname{} to capture changes in program semantics such as control-flow or data-flow to further improve the capability of \toolname{}.
Finally, as the results are promising, \toolname{} can be utilized to automatically curate a benchmark of vulnerabilities that can be used to evaluate vulnerability detection systems. We plan to build a benchmark of vulnerabilities in future work.

\vspace{2mm}

% \noindent \textbf{Data availability.} \toolname{}’s implementation are publicly available at \url{https://github.com/soarsmu/midas}.
% Besides, aside from commit length, studying other aspects, for example the number of context switch, to further increase the number of detected vulnerability-fixing commits under a limited inspection cost is also a worth-investigating direction in the future.
% \bowen{what does the ``other aspects'' refer to? Better give an concrete example.}

% % use section* for acknowledgment
% \ifCLASSOPTIONcompsoc
%   % The Computer Society usually uses the plural form
%   \section*{Acknowledgments}
% \else
%   % regular IEEE prefers the singular form
% \fi


\balance

\bibliographystyle{IEEEtran}
\bibliography{main}

\end{document}